\documentclass{aa}
\usepackage{graphicx}
\usepackage{txfonts}
\usepackage{natbib}
\usepackage{longtable}
\usepackage{multirow}
\usepackage{color}
 \usepackage{url}

\newcommand{\siii}{[\textrm{Si}~\textsc{ii}]}

\newcommand{\nii}{[\textrm{N}~\textsc{ii}]}

\newcommand{\hi}{\textrm{H}~\textsc{i}}
\newcommand{\oiii}{[\textrm{O}~\textsc{iii}]}
\newcommand{\oii}{[\textrm{O}~\textsc{ii}]}
\newcommand{\oiilam}{[\textrm{O}~\textsc{ii}]\ensuremath{\lambda3727}}
\newcommand{\oiiiv}{[\textrm{O}~\textsc{iii}]\ensuremath{\lambda5007}}
\newcommand{\oiiiiv}{[\textrm{O}~\textsc{iii}]\ensuremath{\lambda4959}}
\newcommand{\oiiii}{\textrm{O}~\textsc{iii}]\ensuremath{\lambda1666}}
\newcommand{\oiiidoub}{[\textrm{O}~\textsc{iii}]\ensuremath{\lambda\lambda4959,5007}}
 
\newcommand{\ha}{\ifmmode {\rm H}\alpha \else H$\alpha$\fi}
\newcommand{\hb}{\ifmmode {\rm H}\beta \else H$\beta$\fi}
\newcommand{\lya}{\ifmmode {\rm Ly}\alpha \else Ly$\alpha$\fi}
\newcommand{\lyb}{\ifmmode {\rm Ly}\beta \else Ly$\beta$\fi}
\newcommand{\lyg}{\ifmmode {\rm Ly}\gamma \else Ly$\gamma$\fi}
\newcommand{\dfel}{\ifmmode \Delta\log {\rm f}_{\rm EL} \else $\Delta\log$ f$_{\rm EL}$\fi}
\def\ionii{{\it Ion2}}

\newcommand{\flyc}{\ifmmode  \mathrm{f}_\mathrm{esc}\mathrm{(LyC)} \else $\mathrm{f}_\mathrm{esc}\mathrm{(LyC)}$\fi}
\newcommand{\flya}{\ifmmode  \mathrm{f}_\mathrm{esc}\mathrm{(Ly\alpha)} \else $\mathrm{f}_\mathrm{esc}\mathrm{(Ly\alpha)}$\fi}
\newcommand{\ciii}{\textrm{C}~\textsc{iii}]\ensuremath{\lambda1909}}

\newcommand{\cii}{[\textrm{C}~\textsc{ii}]\ensuremath{\lambda1334}}
\newcommand{\nv}{\textrm{N}~\textsc{v}\ensuremath{\lambda1240}}
\newcommand{\civ}{\textrm{C}~\textsc{iv}\ensuremath{\lambda1550}}
\newcommand{\heii}{\textrm{He}~\textsc{ii}\ensuremath{\lambda1640}}

\def\ergs{\ifmmode \mathrm{erg\hspace{1mm}s}^{-1} \else erg s$^{-1}$\fi}

\def\micron{$\mu$m}
\def\msun{\ifmmode \mathrm{M}_{\odot} \else M$_{\odot}$\fi}
\def\msunyr{\ifmmode \mathrm{M}_{\odot} \hspace{1mm}{\rm yr}^{-1} \else $\mathrm{M}_{\odot}$ yr$^{-1}$\fi}
\def\zsun{\ifmmode Z_{\odot} \else Z$_{\odot}$\fi}
\def\lsun{\ifmmode L_{\odot} \else L$_{\odot}$\fi}
\def\mstar{\ifmmode \mathrm{M}_{\star} \else M$_{\star}$\fi}

\newcommand{\myemail}{stephane.debarros@oabo.inaf.it}

\begin{document}
\title{An extreme \oiii\ emitter at $z=3.2$: a low metallicity Lyman
continuum source}

   \author{S.~de Barros
          \inst{1}
	  \and
          E.~Vanzella
          \inst{1}
         \and
         R.~Amor\'{i}n
          \inst{2}
          \and
          M.~Castellano
          \inst{2}
          \and
          B.~Siana
          \inst{3}
          \and
          A.~Grazian
          \inst{2}
          \and
          H.~Suh
          \inst{4, 8}
          \and
          I.~Balestra
          \inst{5}
          \and
          C.~Vignali
          \inst{1,6}
          \and
          A.~Verhamme
          \inst{7}
          \and
          G.~Zamorani
          \inst{1}
          \and
          M.~Mignoli
          \inst{1}
          \and
          G.~Hasinger
          \inst{8}
          \and
          A.~Comastri
          \inst{1}
          \and
           L.~Pentericci
          \inst{2}
          \and
          E. P\'{e}rez-Montero
          \inst{9}
          \and
          A.~Fontana
           \inst{4}
           \and
           M.~Giavalisco
           \inst{10}
           \and
           R.~Gilli
           \inst{1}
                      }

   \institute{INAF--Osservatorio Astronomico di Bologna, via Ranzani 1, I-40127 Bologna, Italy, email: \myemail
   \and
   INAF--Osservatorio Astronomico di Roma, via Frascati 33, 00040 Monteporzio, Italy
   \and
   Departement of Physics and Astronomy, University of California, Riverside, CA 92507, USA
   \and
   Harvard-Smithsonian Center for Astrophysics, Cambridge, MA 02138, USA
   \and
   INAF--Osservatorio Astronomico di Trieste, via G. B. Tiepolo 11, 34131, Trieste, Italy
   \and
   Dipartimento di Fisica e Astronomia, Universit\`{a} degli Studi di Bologna, Viale Berti Pichat 6/2, 40127 Bologna, Italy
   \and
   Observatoire de Gen\`{e}ve, Universit\'{e} de Gen\`{e}ve, Ch. des Maillettes 51, 1290 Versoix, Switzerland
   \and
   Institute for Astronomy, 2680 Woodlawn Drive, Honolulu, Hawaii 96822, USA
   \and
   Instituto de Astrof\'{i}sica de Andaluc\'{i}a, CSIC, Apartado de correos 3004, 18080 Granada, Spain
   \and
   Astronomy Department, University of Massachusetts, Amherst, MA 01003, USA
   	}

   \date{Received ; accepted}
   
   \authorrunning{} \titlerunning{An extreme \oiii\ emitter at $z=3.212$}
   \abstract
   {}
   {The cosmic reionization is an important process occurring in the early epochs of the Universe. However, because of observational limitations due to the opacity of the intergalactic medium to Lyman continuum photons, the nature of ionizing sources is still not well constrained. 
   While high-redshift star-forming galaxies are thought to be the main contributors to the ionizing background at $z>6$, it is impossible to directly detect  their ionizing emission. Therefore, looking at intermediate redshift analogues ($z\sim2-4$) can provide useful hints about cosmic reionization.}
      {We investigate the physical properties of one of the best Lyman continuum emitter candidate at $z=3.212$ found in the GOODS-S/CANDELS field with photometric coverage from $U$ to MIPS 24\micron\ band and VIMOS/VLT and MOSFIRE/Keck spectroscopy. These observations allow us to derive physical properties such as stellar mass, star-formation rate, age of the stellar population, dust attenuation, metallicity, and ionization parameter, and to determine how these parameters are related to the Lyman continuum emission.}
   {Investigation of the UV spectrum confirms a direct spectroscopic detection of the Lyman continuum emission with $S/N>5$. Non-zero \lya\ flux at the systemic redshift and high Lyman-$\alpha$ escape fraction ($\flya\geq0.78$) suggest a low \hi\ column density. The weak C and Si low-ionization absorption lines are also consistent with a low covering fraction along the line of sight.
The subsolar abundances are consistent with a young and extreme starburst. The \oiiidoub+\hb\ equivalent width (EW) is one of the largest reported for a galaxy at $z>3$ ($\mathrm{EW(\oiiidoub+\hb)} \simeq 1600\AA$, rest-frame; 6700\AA~observed-frame) and the NIR spectrum shows that this is mainly due to an extremely strong \oiii\ emission. The large observed \oiii/\oii\ ratio ($>10$) and high ionization parameter are consistent with prediction from photoionization models in case of a density-bounded nebula scenario. Furthermore, the EW(\oiiidoub+\hb) is comparable to recent measurements reported at $z\sim7-9$, in the reionization epoch.
      We also investigate the possibility of an AGN contribution to explain the ionizing emission but most of the AGN identification diagnostics suggest that stellar emission dominates instead.}
   {This source is currently the first high-$z$ example of a Lyman continuum emitter exhibiting indirect and direct evidences of a Lyman continuum leakage and having physical properties consistent with theoretical expectation from Lyman continuum emission from a density-bounded nebula. A low \hi\ column density, low covering fraction, compact star-formation activity, and a possible interaction/merging of two systems may contribute to the Lyman continuum photon leakage.}
   

   \maketitle
%

   \section{Introduction}
   \label{sec:intro}

One of the most pressing, unanswered questions in cosmology and galaxy
evolution is which are the sources responsible for the
reionization
of the intergalactic medium \citep[IGM;][]{robertsonetal2015} and capable of keeping it
ionized
afterwards \citep{fontanotetal2014,giallongoetal2015}. Until recently, it was difficult to draw a consistent picture
based on the different observational constraints \citep[e.g., reconciling the ionizing background and the galaxy UV luminosity density;][]{bouwensetal2015}.
However, there is still a large number of unconstrained parameters  at $z>6$, the most important one probably being the Lyman continuum escape fraction \citep[\flyc;][]{zackrissonetal2013}.
Unfortunately, it is not possible to directly detect Lyman continuum emission at the epoch of reionization or even in the aftermath (e.g., at any redshift $z> 4.5$), due to the very high cosmic opacity \citep[e.g.,][]{worsecketal2014}.

A number of surveys at $1< z < 3.5$ both from the ground and with the Hubble Space Telescope ({\it HST}), have looked for ionizing photons by means of imaging or spectroscopy and there have been some claims of detections \citep{steideletal2001,shapleyetal2006,nestoretal2013,mostardietal2013,mostardietal2015}. However, careful analysis of some claims combining different instruments \citep[e.g., near-infrared spectroscopy and high spatial resolution probing LyC,][]{sianaetal2015} shows that the individual detections must be considered with caution because of the difficulty to eliminate a possible foreground contamination \citep{vanzellaetal2010b,mostardietal2015}. Several other surveys reported only upper limits \citep{sianaetal2010,bridgeetal2010,malkanetal2003,vanzellaetal2010b,V12,boutsiaetal2011,grazianetal2015}.
This could be due to the rarity of relatively bright
ionizing sources, as a consequence of the combination of view-angle effects
\citep{cen&kimm2015}, stochastic intergalactic opacity
\citep{inoue&iwata2008,inoueetal2014} and possibly intrinsically low escaping ionizing
radiation on
average, in the luminosity regime explored so far \citep[$L > 0.1L^{*}$, e.g.,][]{vanzellaetal2010b}.

At higher redshifts, the galaxy contribution to the cosmic ionization budget must be
inferred from
other properties that correlate with their ability to contribute to
ionizing radiation,
if such properties exist.
This is an open line of research with ground and space-based facilities.
Indirect signatures in the non-ionizing domain have been investigated by
performing photoionization and radiative transfer modeling. Low column
density
and covering fraction of neutral gas (that regulate the escape fraction of
ionizing
radiation) are related to a series of possible indicators:
\begin{enumerate}
\item the nebular conditions, (i.e. density-bounded or radiation bounded) traced
by peculiar line
ratios like large \oiii/\oii\ ratio \citep[e.g.,][]{jaskot&oey2014,nakajima&ouchi2014} or
deficit of Balmer emission lines given the inferred star-formation rate (SFR) and the UV $\beta$ slope
\citep[$F_\lambda\propto\lambda^\beta$;][]{zackrissonetal2013};
\item strength of interstellar absorption lines as a trace of neutral gas
covering fraction \citep{jonesetal2012,heckmanetal2011};
\item structure of the \lya\ emission line, its width and position
relative to
the systemic redshift \citep[e.g.,][]{verhammeetal2015};
\item morphology and spatial distribution of the star formation
activity (e.g., nucleation) which seems to be another property that favors efficient
feedback and eventually a transparent medium for ionizing radiation \citep[e.g.,][]{heckmanetal2011,borthakuretal2014}.
\end{enumerate}
While all these features are promising signatures, possibly usable at the redshifts of the
reionization, there is still a missing direct link between them and
the Lyman continuum emission. In this respect there is an ongoing effort to
investigate this possible link in a particular class of sources that seem to possess all the above
properties. The idea is to look for ``green pea" (GP) galaxies in the
nearby Universe. GPs have been discovered in the Galaxy Zoo project \citep{cardamoneetal2009}. They are small compact galaxies in the SDSS survey at $0.11<z<0.36$, dominated by strong \oiiidoub\ emission ($\mathrm{EW(\oiiidoub)}>100$\AA, up to $\sim1000$\AA). This strong emission is thought to be related to a non zero ionizing photon escape fraction \citep[e.g,][]{jaskot&oey2014,nakajima&ouchi2014}.
The typical physical properties of GPs are a low stellar mass ($\sim10^{8.5}\msun< \mstar< 10^{10.5}\msun$), low metallicity \citep[$12+\log(\mathrm{O}/\mathrm{H}) = 8.05\pm0.14$,][]{amorinetal2010,amorinetal2012}, high sSFR \citep[sSFR=SFR/stellar mass, 1 to 10 Gyr$^{-1}$,][]{cardamoneetal2009}, young age (few Myr), showing Wolf-Rayet signatures (e.g., bumps and $\textrm{He}~\textsc{ii}]\ensuremath{\lambda4686}$ emission line),
and extremely large \ha\ and \oiiiv\ equivalent widths. Their strong \lya\ emission has been recently observed \citep{henryetal2015}.
However, currently, no direct LyC signatures have been identified in green peas and the work is still ongoing.

Similarly, it would be desirable to find Lyman continuum emitters at even higher redshift, within the first 2 Gyr after the
Big-Bang ($z>3$) and to look for properties and physical mechanisms that allow ionizing photons to escape. \cite{amorinetal2014b} report the properties of a $z=3.417$ galaxy with possible indirect signature of ionizing photon leaking such as a large \oiii/\oii\ ratio, but only an upper limit on the Lyman continuum emission.
Very recently  a class of candidate of strong \hb+\oiii\
emitters has been identified at much higher redshift ($z>6.5$) by looking
at the {\it Spitzer}/IRAC excess in channels 1 or 2
\citep[depending on redshift;][]{schaerer&debarros2010,onoetal2012,labbeetal2013,smitetal2014,smitetal2015,robertsborsanietal2015}.
The redshift of two of them have been spectroscopically confirmed at $z=7.73$ and 8.67 with a combined equivalent width $\mathrm{EW(\hb+\oiii)}\simeq720$\AA\ and $>1000$\AA\ (rest-frame), respectively \citep{oeschetal2015,robertsborsanietal2015,zitrinetal2015}.

The nature of these objects is not well understood and further investigation is needed, though this will have to await the James Webb Space Telescope and the Extremely Large Telescopes.

We have identified two Lyman continuum leakers in \cite{V15}, hereafter V15: {\it Ion1} and \ionii. 
These galaxies have been selected as Lyman continuum emitters through a photometric selection which is based on the
comparison between the observed photometric fluxes and colors probing the Lyman continuum emission and predictions from the combination of spectral synthesis models \cite[e.g.,][]{BC03} and intergalactic medium (IGM) transmissions \citep{inoueetal2014}.

In this work we present the source
with one of the largest \oiiidoub\ line equivalent width
currently known ($\mathrm{EW}\sim1500\AA$, rest-frame) at $z>3$ and with a plausible
leakage of ionizing radiation.
Ultraviolet (VLT/VIMOS), optical and NIR (Keck/MOSFIRE) rest-frame
spectroscopy are presented, as well as a detailed multi-frequency analysis
with the aim to test the aforementioned signatures of linking Lyman continuum.
We also use the
extreme line ratios to investigate the nebular conditions.
To this aim we also compare the observed \ionii\ properties
with theoretical predictions from photoionization models.

\ionii\ shares some of the properties of sources recently
identified
at much higher redshift (in particular the large EW(\hb+\oiii) and is therefore relevant to reionization,
with the advantage to anticipate
the rest-frame optical spectroscopic investigation of such extreme objects.

The paper is structured as follows. The spectroscopic and photometric data are described in Sect.~\ref{sec:data}. In Sect.~\ref{sec:phys}, we present the physical properties derived from the reanalyzed UV spectrum and the newly acquired NIR MOSFIRE spectrum, and alongside we present the properties derived from the photometry. We also review previously reported properties (V15) in the light of the new data.
In Sect.~\ref{sec:agn}, we discuss a possible AGN contribution to the observed \ionii\ spectra. 
We summarize our main conclusion in Sect.~\ref{sec:conclu}.

    \begin{figure*}[htbf]
  \centering
  \includegraphics[width=18cm,trim=0cm 0cm 0cm 0cm,clip=true]{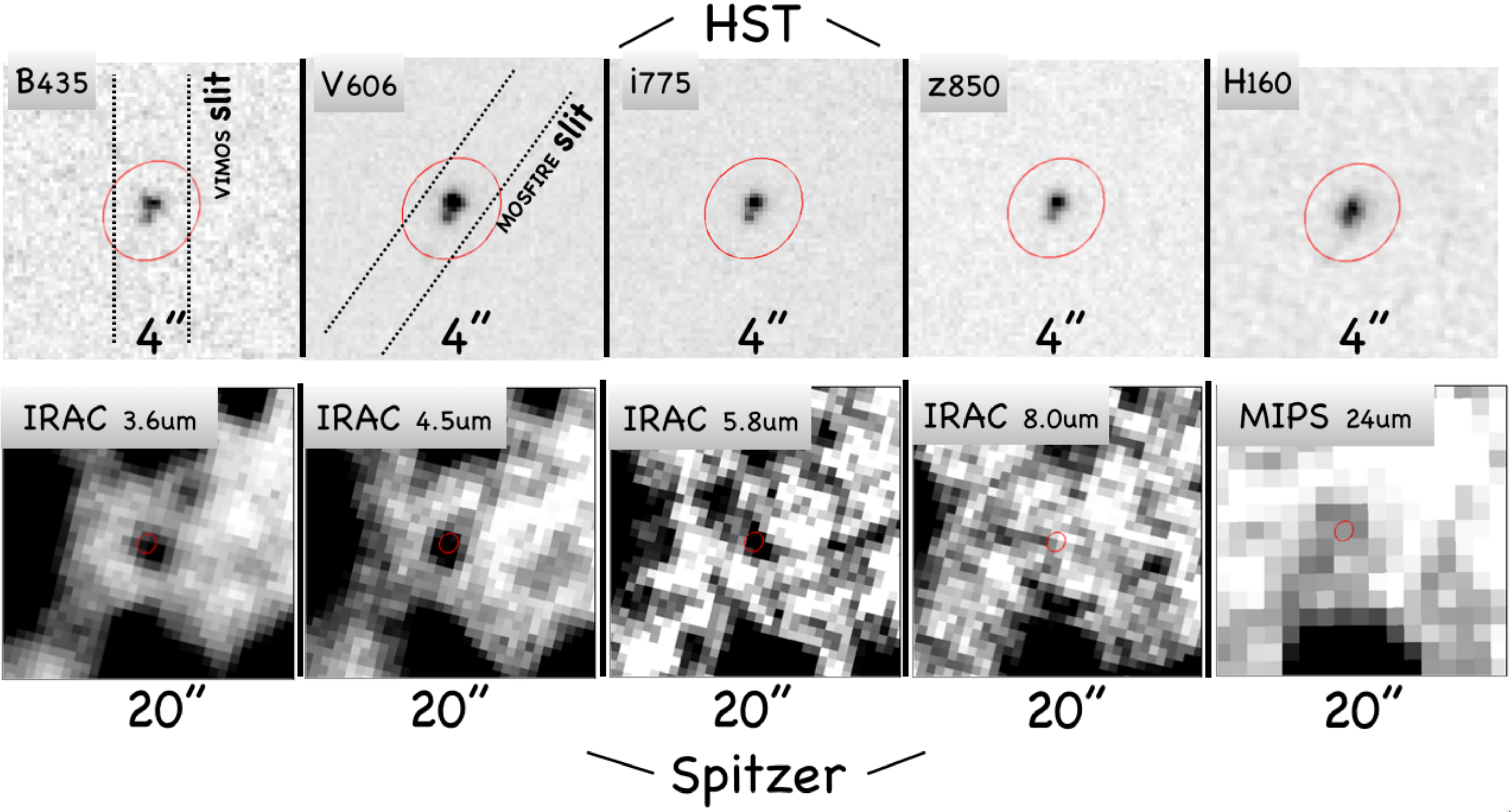}
    \caption{Postage stamps of \ionii\ in $B_{435}$, $V_{606}$, $i_{775}$, $z_{850}$, $H_{160}$, IRAC 3.5\micron, IRAC 4.5\micron, IRAC 5.8\micron, IRAC 8.0\micron\ and MIPS 24\micron. The sizes
    of the images are $4''\times4''$ for the {\it HST} bands and $20''\times20''$ for the {\it Spitzer} bands.}
  \label{fig:fig1}
\end{figure*}

We adopt a $\Lambda$-CDM
cosmological model with $\mathrm{H}_0=70$ km s$^{-1}$ Mpc$^{-1}$,
$\Omega_{\mathrm{m}}=0.3$ and $\Omega_{\Lambda}=0.7$. We assume a
Salpeter IMF \citep{salpeter1955}. All magnitudes are expressed in the
AB system \citep{oke&gunn1983}.


  \section{Data}
  \label{sec:data}
  
  \subsection{Spectroscopy}
  
The available low (LR, R=180) and medium (MR, R=580) resolution VLT/VIMOS spectroscopy
covering the spectral ranges 3400-7000\AA~and 4800-10000\AA~have
been obtained during 2005, with 4hr exposure time for both LR and MR spectra.
The LR spectrum continuum sensitivity is  $S/N\sim0.8$ at 3700\AA\ with 4 hours and m(AB)$=26.5$
assuming seeing 0.8'' and airmass 1.2. It is calculated over one pixels along dispersion and 2xFWHM along spatial direction. \citep{popessoetal2009,balestraetal2010}.
Slit widths of $1''$ were adopted and the median seeing during observations
was $0.8''$.
We have newly reduced both configurations, in particular focusing on the
Lyman continuum emission (LR-spectrum) and the ultraviolet spectral
properties in the non-ionizing domain (MR-spectrum).
Reduction has been done as described in \cite{balestraetal2010}
as well as adopting the AB-BA method described in \cite{vanzellaetal2011,vanzellaetal2014}.
Both methods produce consistent results.

Recently, a near infrared spectrum of \ionii\ has also been obtained with
Keck/MOSFIRE during spring 2015 (PI: G\"{u}nther Hasinger). 32, 32 and 36 minutes integration time
have been obtained in the $J$, $H$ and $K$ bands, respectively, with the AB-BA dithering pattern of 2.9$''$. The adopted slit width was 0.7$''$
producing a dispersion of 1.30\AA, 1.63\AA, 2.17\AA~per pixel, in the
J, H and K band, respectively; the pixels scale along the spatial 
direction was 0.18$''$/pix. \ionii\ was inserted in a mask  aimed to target also low-$z$ objects. The $J$ band is useful to monitor low-$z$ emission lines.

Reduction has been performed using the
standard pipeline\footnote{\url{https://code.google.com/p/mosfire/}}.
Briefly, two-dimensional sky-subtracted spectra, rectified and
wavelength calibrated are produced, from which the one dimensional
variance weighted spectrum is extracted with a top-hat aperture.
Wavelength solution has been computed from sky emission lines \citep[e.g.,][]{rousselotetal2000} in the
three bands
reaching an RMS accuracy better than 1/20, 1/30 and 1/40 of the pixel size in $J$, $H$ and $K$ bands, respectively (typical r.m.s. $< 0.05$\AA).

      \begin{figure*}[htbf]
  \centering
  \includegraphics[width=18cm,trim=0cm 0cm 0cm 0cm,clip=true]{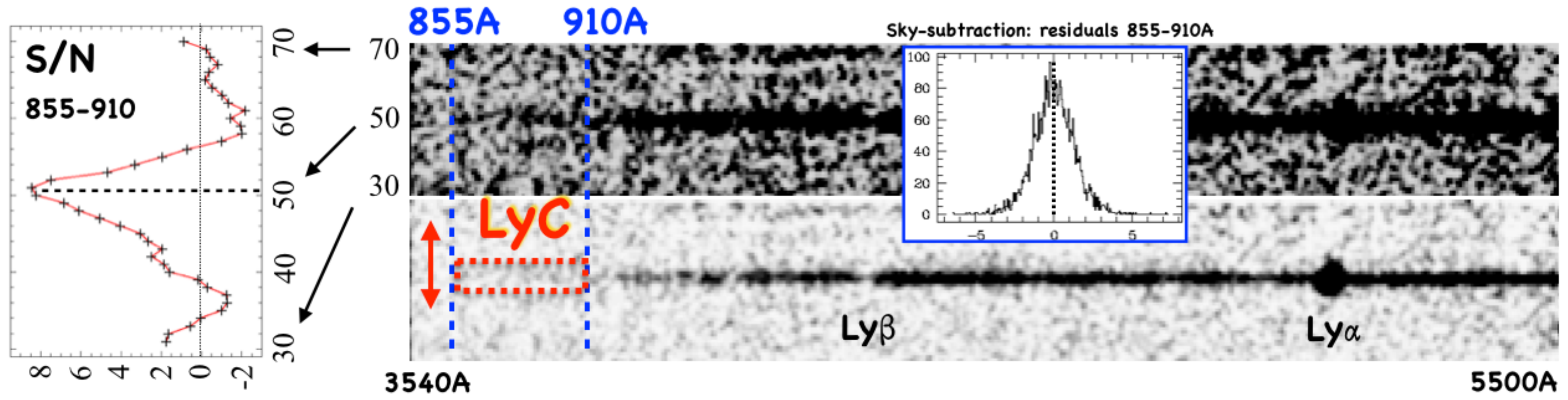}
    \caption{Two-dimensional LR VIMOS UV spectrum of \ionii\  with different cuts to highlight the spectral features \lya\ and \lyb\ (bottom) and the Lyman continuum $\lambda<912\AA$, in the range 855-910\AA\ (top, vertical dotted blue lines). We show the moving average calculated within a rectangular aperture (855-910\AA\ $\times$ 1.25'', red-dotted rectangle) in the spatial direction divided by its r.m.s.  on the left side. A signal is detected at $\lambda<912\AA$ with $S/N>5$. The inset shows the pixel distribution of the background after sky-subtraction in the region 855-910\AA\ (derived from the $S/N$ spectrum).
The distribution is symmetric ( skewness = -0.016) with the mean and median close to zero, +0.0057 and -0.014, respectively. No significant systematic effects are present.
}
  \label{fig:fig2}
\end{figure*}

Flux calibration has been performed by observing the
standard star A0V HIP 26577. Given the slit width ($0.7''$)  and the
median seeing conditions during observations ($0.85''$), particular care must be
devoted to slit losses. For this reason we check it by using a star
included in the same science mask (WISE 895), for which the continuum is very well detected in the entire MOSFIRE wavelength range, and whose $J$, $H$ and $K$ band total magnitudes are known from CANDELS \citep{groginetal2011,koekemoeretal2011} and HUGS photometry \citep{fontanaetal2014} with typical $S/N > 40$.
From the star WISE 895 and its near infrared photometry 
we derive a correction factor for slit losses and possible centering effects of about 50\%.
With this correction the total line flux (\hb+\oiiidoub) derived from the MOSFIRE spectrum is fully consistent with the total line flux derived from the SED fit and 
$K$-band excess (see Section~\ref{sec:photprop}).
Ion2 is composed by two blobs separated by $0.2''$ (see Figure~\ref{fig:fig1}\footnote{Created with the Rainbow Navigator tool, \url{https://rainbow.fis.ucm.es/Rainbow_navigator_public/}}) each one with a half-light 
radius $Re \simeq 0.3$kpc, or less than $0.1''$ at $z=3.212$. Therefore the
slit losses for the entire system at the $0.7''$ MOSFIRE entrance slit are modulated primarily by the 
seeing during observations, which was on average $0.85''$.
A correction factor of 1.4 is consistent with the expected fraction of light
falling outside the slit for a perfect centering of the target and Gaussian seeing with FWHM=$0.85''$. 
The correction is also compatible with the discussion in \cite{steideletal2014}. Therefore a correction factor of 1.5 is reasonable in our configuration, observational conditions and including
a possible small centering uncertainty.

Line fluxes, errors and limits from VIMOS and MOSFIRE spectra are
reported in Table~\ref{tab:data}.
   

  \subsection{Photometry}
  
Deep $U$ band imaging has been taken with the VLT/VIMOS instrument providing a 1-$\sigma$ depth of mag $\sim 30$ within 1 circular aperture. We refer the reader
to \cite{noninoetal2009} for a complete description of the data reduction \citep[see also][]{vanzellaetal2010c}.
  
          \begin{table*}[htbf]
         \centering
         \caption{Emission line properties.}
         \begin{tabular}{lccccc}
           \hline
           Line & Flux$^{\mathrm{a}}$ & Dust corrected flux$^{\mathrm{b}}$ &FWHM$^{\mathrm{c}}$ & EW$^{\mathrm{d}}$ & $z$\\
           \hline
	   \lya\            &  $15.4\pm1.0^{\mathrm{e}}$  & $15.4^{+26.0}_{-1.0}$ & $240\pm60^{\mathrm{f}}$  &  $94\pm20^{\mathrm{e}}$ & $3.2183^{\mathrm{f}}\pm0.003$ \\
           \lya\ (blue)  &  $5.1$                                     & \ldots & \ldots                                  &  \ldots                                & $3.2126\pm0.005$ \\
           \lya\ (red)    &  $10.3$                                   &  \ldots & \ldots                                  &  \ldots                                & $3.2183\pm0.003$ \\
           \nv\             &  $<0.16$                                 & $<0.4$ & \ldots                                   &  \ldots                                & \ldots \\
           \civ\            &  $<0.18$                                  & $<0.4$ & \ldots                                   &  \ldots                                 & \ldots \\
           \heii$^{\mathrm{g}}$  & $0.6\pm0.3$            & $0.6^{+1.3}_{-0.3}$ & \ldots                                   & \ldots                                 & \dots \\
           \oiiii$ ^{\mathrm{g}}$ & $0.5\pm0.3$             & $0.5^{+1.2}_{-0.3}$ & \ldots                                   & \ldots                                 & \dots \\
           \ciii\            &  $1.3\pm0.3$                          & $1.3^{+2.3}_{-0.3}$ & $400\pm90^{\mathrm{h}}$  &  $18^{+9}_{-5}$                 & $3.2127$\\
           \oiilam$^{\mathrm{i}}$ &  $1.5\pm1.5$          & $1.5^{+3.3}_{-1.5}$ & \ldots                                    &  $<50$                              &  $3.2126$  \\
           \hb$^{\mathrm{g}}$ &  $1.5\pm0.8$               & $1.5^{+1.5}_{-0.8}$ & \ldots                                    & $112\pm60$                      &  $3.2128\pm0.005$  \\
           \oiiiiv\         &  $5.4\pm0.8$ $(7.4\pm1.1)$     & $7.4^{+4.6}_{-1.1}$ & $155\pm70$                        & $365\pm70$                      &  $3.2128\pm0.0008$  \\
           \oiiiv\         &  $22.1\pm0.8$                            & $22.1^{+10.8}_{-0.8}$ & $147\pm30$                        & $1103\pm60$                    &  $3.2126\pm0.0006$  \\
           \hline
         \end{tabular}
         \label{tab:data}
         \begin{list}{}{}
         \item[$^{\mathrm{a}}$ Observed emission line flux and error corrected for slit losses (50\%)  in units of $10^{-17}$ erg s$^{-1}$ cm$^{-2}$. In parenthesis, the \oiiiiv\ flux]
         \item[adopting the theoretical ratio $\oiiiv/\oiiiiv = 3$ \citep{storey&zeippen2000}.]
         \item[$^{\mathrm{b}}$ The emission line dust correction method is described in Appendix~\ref{ap:el}.]
         \item[$^{\mathrm{c}}$ FWHM (instrumental corrected) and error in units of km s$^{-1}$.]
         \item[$^{\mathrm{d}}$ Equivalent width and error in units of \AA.]
         \item[$^{\mathrm{e}}$ Sum of the two components.]
         \item[$^{\mathrm{f}}$ Main component.]
         \item[$^{\mathrm{g}}$ Marginal detections (Figure~\ref{fig:fig6}): $S/N(\oiiii)\sim1.7$, $S/N(\heii)\sim2$, and $S/N(\hb)\sim2$]
         \item[$^{\mathrm{h}}$ FWHM is the contribution of the two blended components, 1907\AA-1909\AA.]
         \item[$^{\mathrm{i}}$ The error is large because the line is on the top of a sky line (Figure~\ref{fig:fig4}).]
         \end{list}
      \end{table*}
  
The photometry from the $U$ VIMOS band to the {\it Spizter}/IRAC 8.0\micron\ band is taken from the CANDELS GOODS-S
multi-wavelength catalog \citep{guoetal2013}. The optical {\it Hubble Space Telescope} ({\it HST}) images from the Advanced Camera for Surveys (ACS) combine the data from \cite{giavaliscoetal2004b}, \cite{beckwithetal2006} and \cite{koekemoeretal2011}. The field was observed in the F435W, F606W, F775W, F814W and F850LP bands. Near-infrared WFC3/IR data combines data from \cite{groginetal2011}, \cite{koekemoeretal2011} and \cite{windhorstetal2011}, with observations made in the F098M, F125W, 
and F160W bands. The VLT/HAWK-I $K_s$ band imaging are taken from the HUGS survey \citep{fontanaetal2014}. The {\it Spitzer}/IRAC 1 and 2 data are taken from the SCANDELS survey \citep{ashbyetal2015}, while the IRAC 3 and 4 are taken from \cite{ashbyetal2013}. Hereafter, we will refer to the filters F435W, F606W, F775W, F814W, F850LP, F098M, F125W, 
F160W as $B_{435}$, $V_{606}$, $i_{775}$, $I_{814}$, $z_{850}$, $Y_{098}$, $J_{125}$, $H_{160}$, respectively.

We also check detection in {\it Spitzer}/MIPS bands and in the {\it Herschel} data. As shown in Figure~\ref{fig:fig1}, we are able to put an upper limit to the MIPS 24\micron\ flux \citep[$S/N<2$, $m_\mathrm{AB}=22.27\pm0.90$,][]{santinietal2009}, while it is undetected in {\it Herschel} bands ($<1.2$mJy, $<1.2$mJy, $<2.4$mJy in the 70\micron\, 100\micron, and 160\micron\ bands, respectively, 5$\sigma$ upper limits).


  \section{Physical properties}
  \label{sec:phys}
  
\subsection{Spectroscopic detection and spatial distribution of the Lyman continuum emission}
\label{sec:lyc}

    \begin{figure}[htbf]
  \centering
  \includegraphics[width=9cm,trim=0cm 0cm 0cm 0cm,clip=true]{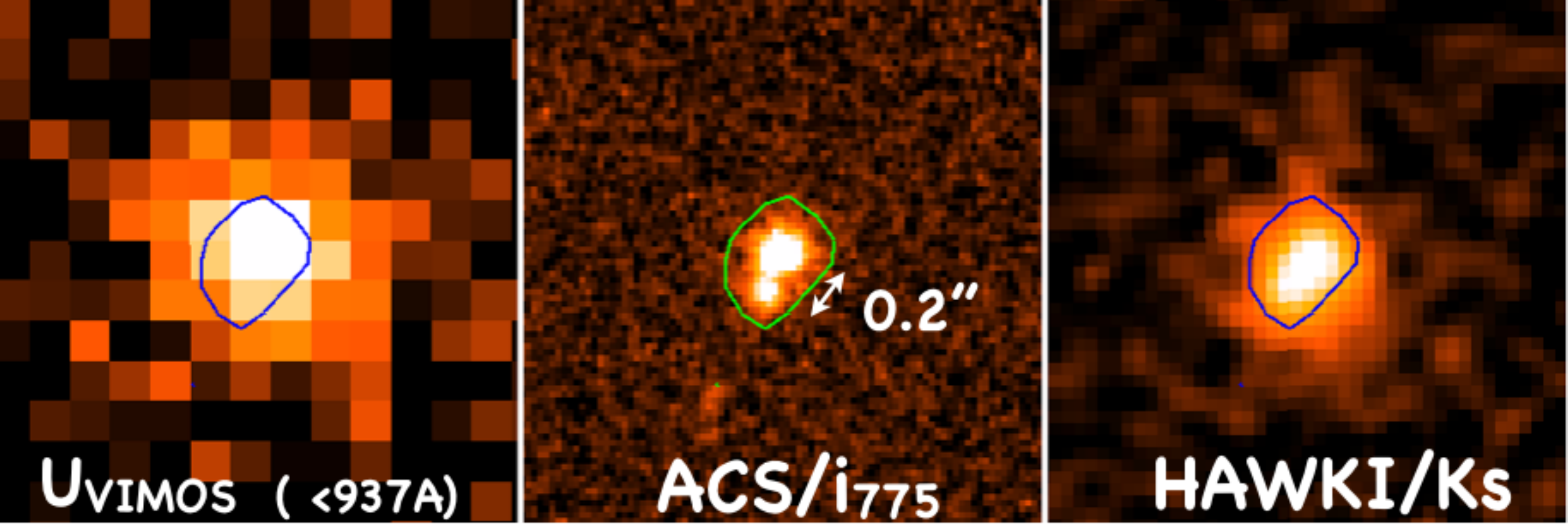}
    \caption{Ground based VLT/VIMOS U-band observation of \ionii\ at a resolution of 0.2''/pixel (left panel). The contour of the system (derived from the ACS/$i_{775}$ band, center panel) is indicated and superimposed in blue to the VIMOS U-image and to the HAWK-I $K_s$ image (right panel). The cutouts are 2.6" vs. 2.6".}
  \label{fig:fig3}
\end{figure}

Spectroscopic detection of stellar Lyman continuum emission at $z\sim3$ has been previously claimed  for few galaxies \citep[e.g.,][]{steideletal2001,shapleyetal2006} but this kind of detection remains a difficult task, mainly because of foreground contamination \citep{vanzellaetal2010b,sianaetal2015,mostardietal2015}, and a clear spectroscopic detection of Lyman continuum emission  at high-redshift is still missing. The only two spectroscopic detections reported in literature are from \cite{steideletal2001} in a stack of 29 LBGs and individually in a subsequent spectroscopic follow-up, in 2 out of 14 LBGs observed by \cite{shapleyetal2006}, dubbed D3 and C49. \cite{vanzellaetal2010b} performed dedicated simulations to derive the probability of
foreground contamination, specifically for \cite{steideletal2001} and \citep{shapleyetal2006} adopted observational setups. In both cases, the probability was high enough to rise the doubt on their reliability. Subsequently the two sources with possible LyC detection of \cite{shapleyetal2006} have not been confirmed as LyC sources. In particular D3 was ruled out with much deeper narrow band imaging by
\cite{iwataetal2009} and \cite{nestoretal2011} and C49 was excluded from the list of LyC sources after the possible spectroscopic confirmation of a close foreground lower redshift source \citep{nestoretal2013} and finally confirmed to be a z=2.029 foreground source \citep{sianaetal2015}. Therefore, at present there are not direct LyC detections from star-forming galaxies, apart the source reported in the present work.

The VIMOS LR spectrum of
\ionii shows a clear signal detected blueward the Lyman limit. We performed a careful reanalysis of the \ionii\ low-resolution UV spectrum by computing the moving average of the flux at $\lambda<910\AA$ and we find a signal with $S/N>5$ (Figure~\ref{fig:fig2}). While several systematics can affect the derivation of the LyC signal \citep[e.g., background substraction][]{shapleyetal2006}, we use different methods of sky subtraction
(ABBA method and polynomial fitting with different orders) and they all provide consistent results.
In particular the same moving average in the LyC region
calculated over the $S/N$ two-dimensional spectrum we derived with ABBA method \citep[see][]{vanzellaetal2014b}
produces stable results with no significant systematics and a signal ($>5\sigma$)
at the same spatial position of the target. This signal can be interpreted as a direct detection of the Lyman continuum emission, 
but the presence of the two components in the {\it HST} images
can cast some doubts about the origin of this detection. However, assuming that the faintest component keeps the same magnitude as the derived $B_{435}$ magnitude ($27.25\pm0.24$; V15) at shorter wavelengths, the average $S/N$ in the range 3600-3840\AA\ is $\sim0.5$ per pixel (from ETC). Averaging over 20 pixels, as we did in Figure~\ref{fig:fig2}, we would expect $S/N\sim2.2$. In addition, the $B$-band dropout signature of such a component ($B-V=0.62$, V15) between the $B_{435}$ and $V_{606}$ 
prevents a possible increased emission approaching the $U$-band, unless an emission feature is boosting the $U$-band magnitude. In such a case the only possible line would be \oiilam\ which would also imply a certain amount of star-formation activity, that in turns 
would be detectable through Balmer and/or Oxygen lines in the wide spectral range probed here. 
The most likely explanation is that the spectroscopic detection is due to a Lyman continuum emission emerging from the brightest \ionii\ component. Furthermore, the ground-based VLT/VIMOS $U$-band spatial distribution shows that most of the $U$-band flux is emitted from the brightest component (Figure~\ref{fig:fig3}).

However, the $U$-band probes both ionizing {\it and} non-ionizing photons ($\lambda<937\AA$), so a fraction of the signal is not due only to Lyman continuum photons, and the ground-based observations are clearly limited in terms of resolution, as seen in Figure~\ref{fig:fig3}. 

The only way to clarify the exact position and the detailed spatial distribution of the Lyman continuum emission is to perform dedicated {\it HST} observations. A proposal to observed \ionii\ with {\it HST}/F336W (17 orbits) has been recently approved (PI: Vanzella, cycle 23).
Hopefully, emission line diagnostics can be used to characterize the gaseous and stellar content of \ionii\ and provide some hints about a Lyman continuum leakage, as discussed in the next Section.

From the signal detected at $880-910\AA$, we derive the relative \flyc\ defined as:
\begin{eqnarray}
\label{eq:deffrel}
f_{esc,rel} = \frac{(L_{UV}/L_{LyC})^{intrinsic}}{(F_{UV}/F_{LyC})^{observed}}\exp(\tau_{LyC}),
\end{eqnarray}
where $(L_{UV}/L_{LyC})^{intrinsic}$ is the intrinsic UV to ionizing luminosity density, $(F_{UV}/F_{LyC})^{observed}$ is the observed UV to ionizing flux density, and $\exp(\tau_{LyC})$ is the LyC attenuation due to the IGM. The signal measured in the LR spectum corresponds to $\sim26.95$ AB magnitude and $m_{1500}=24.60$ is the magnitude observed at 1500\AA\ rest-frame (using the $V_{606}$ band corrected for the \lya\ flux).
The average IGM transmission is 0.54 in the $880-910\AA$\ range \citep{inoue&iwata2008,inoueetal2014}  and assuming $L_{1500}/L_{900}=3$ \citep{nestoretal2013}, we get:
\begin{eqnarray}
\label{eq:frel}
f_{esc,rel}=0.64^{+1.10}_{-0.10}
\end{eqnarray}
If we relax the assumption about the intrinsic $L_{1500}/L_{900}$ ratio, then $f_{esc,rel}$ increases. Or conversely, assuming the maximum IGM transmission at $z=3.2$ (0.74), $L_{1500}/L_{900}$ must be lower than 6.5 to keep $f_{esc,rel}<1$.


\subsection{Physical properties of the interstellar medium}
\label{sec:elprop}

As described in Section~\ref{sec:data}, we use VLT/VIMOS and Keck/MOSFIRE spectroscopy to obtain a wide \ionii\ spectral coverage with $850\AA<\lambda<5700\AA$ (rest-frame). The emission line fluxes, full widths half maximum and equivalent widths are summarized in Table~\ref{tab:data}. Several lines are barely detected with signal-to-noise $< 3$ (e.g., \hb). However, thanks to the well constrained redshift (see below) and because we know exactly the wavelength position of
some features, we do not expect strong systematics.
This is the case for the \hb\ and \oiilam\ lines, whose signature are indeed at the expected positions.

 \begin{figure*}[htbf]
  \centering
  \includegraphics[width=18cm,trim=0.25cm 0cm 0cm 0cm,clip=true]{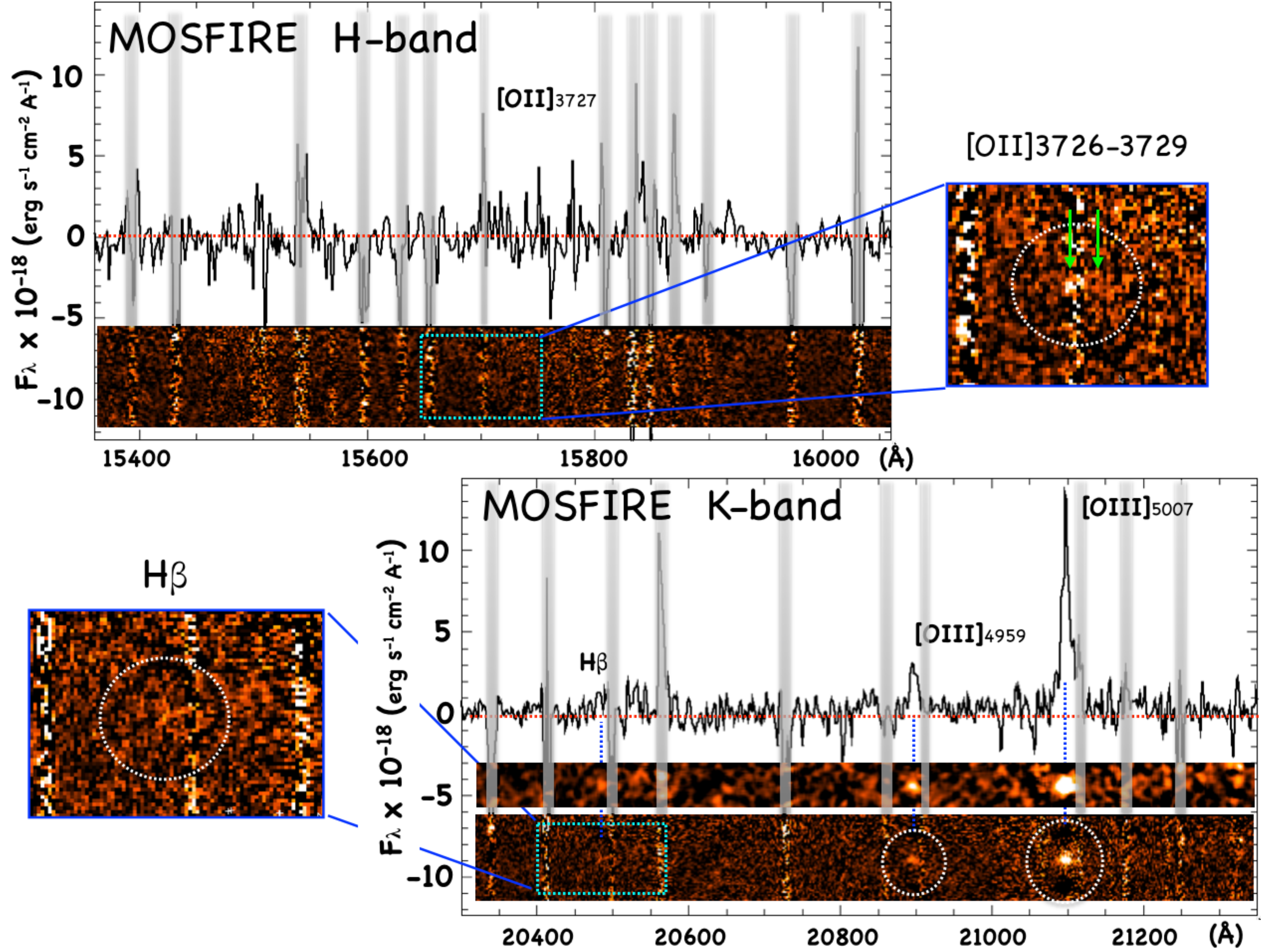}
    \caption{Two- and one-dimensional NIR spectra of \ionii\ in the MOSFIRE $H$ and $K$-band. Two insets show the regions where \hb\ and the \oiilam\ doublet are expected accordingly with the redshift. The gray stripes in the one-dimensional spectra indicate the position of the sky lines.}
  \label{fig:fig4}
\end{figure*}

\begin{figure*}[htbf]
  \centering
  \includegraphics[width=18cm,trim=1cm 0cm 0.8cm 0cm,clip=true]{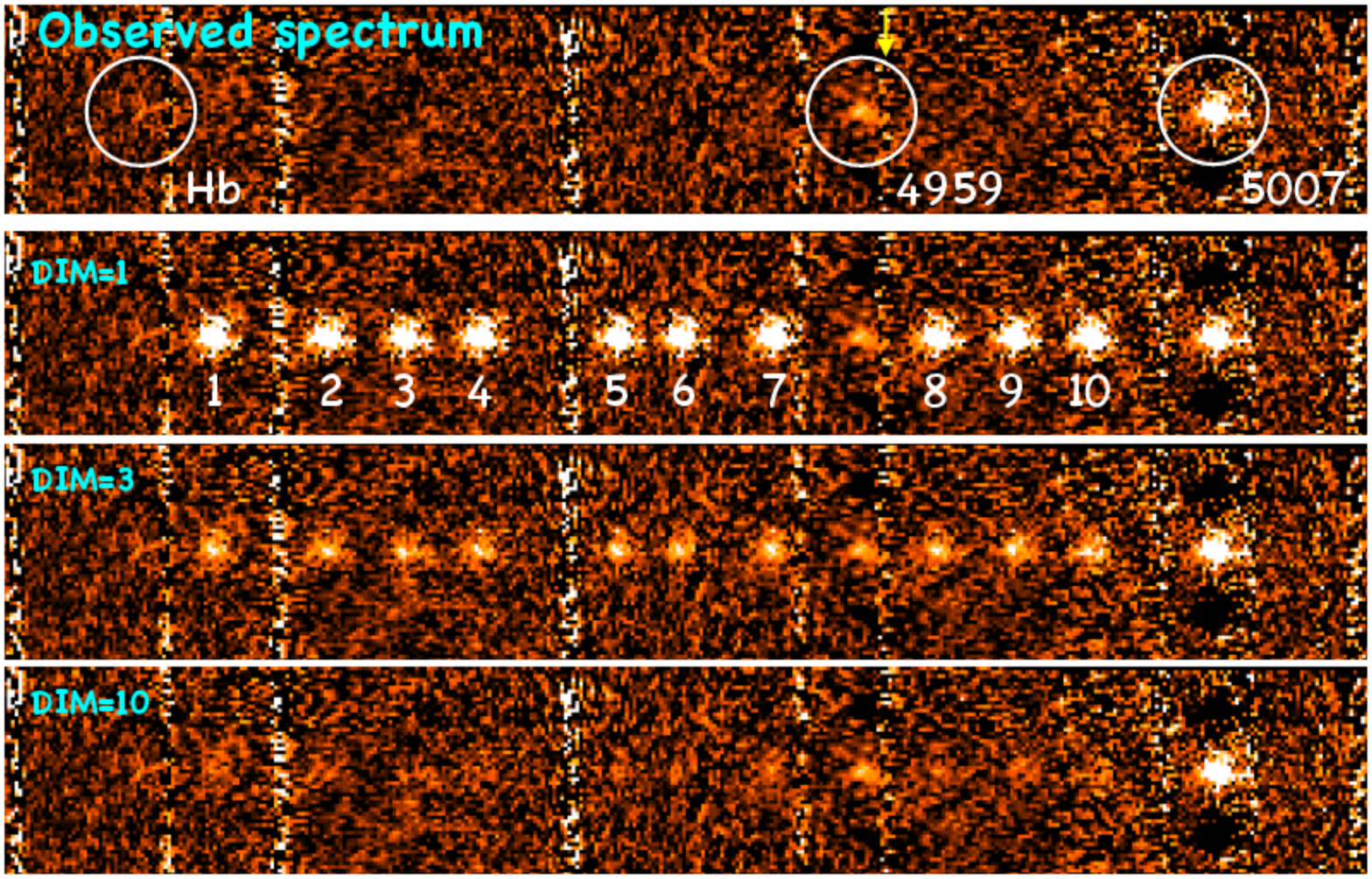}
    \caption{Observed and simulated two-dimensional MOSFIRE spectrum, with the observed \oiiiv\ line replicated at different wavelength positions with different dimming factors (DIM). See text.}
  \label{fig:fig5}
\end{figure*}

Individual line flux errors have been derived using the error spectrum. Additionally, we perform simple simulations to test the error associated to emission lines ratios in the MOSFIRE spectrum (Figure~\ref{fig:fig4}). Figure~\ref{fig:fig5} shows the positions where the \oiiiv\ observed line has been added to the sky-subtracted spectrum (marked with numbers from 1 to 10) with different dimming factors (DIM=1, 3, 10, from top to bottom). This provides  a rough estimate of the typical error and $S/N$ ratio of our measurements. For example, the typical error associated to the line ratio is $\oiiiv/\oiiiiv=4.1\pm0.7$, which is consistent with the theoretical ratio (2.99) within 1.4-$\sigma$ (if necessary, we fix the ratio to the theoretical one). Furthermore, the \oiiiv\ line dimmed by a factor 10 (lowest panel in Figure~\ref{fig:fig5}) is similar to the observed \hb\ 2D spectrum, indicating a likely \oiiiv\ to \hb\ ratio $\geq10$. The marginal \hb\ detection (with $S/N\sim2$) allows us to derive a ratio $\oiii/\hb = 14.7 \pm 5.1$, while for the \oii\ line the non-detection translates to a 2-$\sigma$ lower limit of $\oiiiv/\oiilam>10$ (uncorrected for dust in both cases). The errors derived with our simulations are fully consistent with  those derived from the error spectrum.

\begin{figure*}[htbf]
  \centering
  \includegraphics[width=18cm,trim=0cm 0cm 0cm 0cm,clip=true]{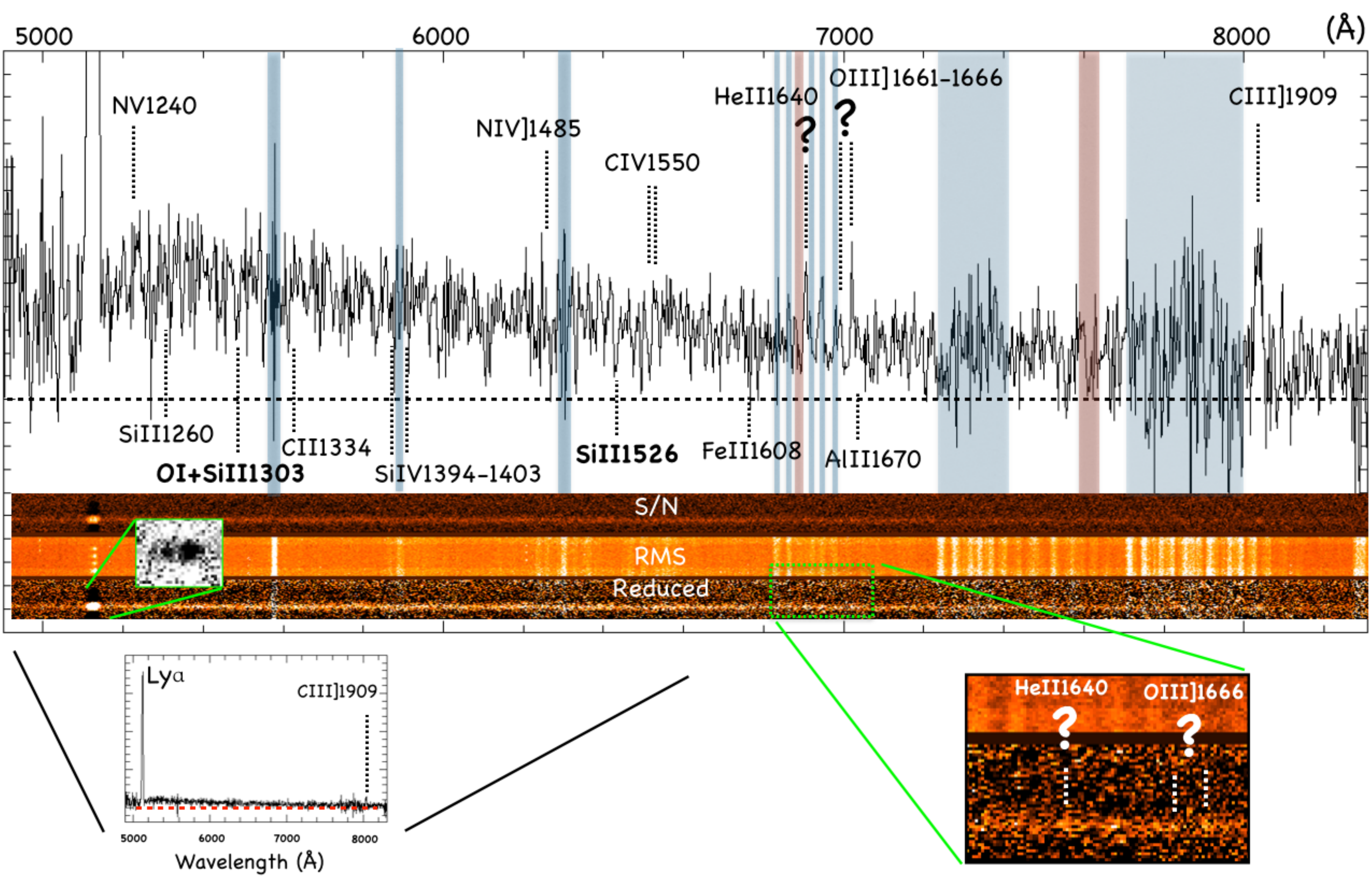}
    \caption{1D and 2D UV VLT/VIMOS MR spectra. The  transparent blue stripes indicate position of the emission sky lines. The transparent red stripes indicate the telluric molecular sky absorptions bands, $B$ and $A$ bands at $\sim6900\AA$ and $\sim7600\AA$. In the same panel, three two-dimensional spectra are shown, the reduced, r.m.s. and signal to noise, from bottom to top, respectively. The inset on the left side shows the double peaked Lya line (from V15). In the bottom-right a zoom in around \heii\ and \oiiii\ positions is shown. Their detection remain tentative. In the bottom-left, the 1D spectrum is shown allowing to see the relative \lya\ flux in comparison with other lines.}
  \label{fig:fig6}
\end{figure*}

The first property derived from emission lines is the spectroscopic redshift. As already reported in V15, the $\textrm{C}~\textsc{iii}]\ensuremath{\lambda1906.68-1908.68}$ transition is clearly detected in the VIMOS MR spectrum (with $S/N=8$, Figure~\ref{fig:fig6}). This feature shows a symmetric shape with a relatively large FWHM ($=400 \mathrm{km\hspace{1mm}s}^{-1}$) with respect to other lines like \lya\ and \oiii, suggesting the two components have similar intensities, even if they are blended and non resolved.  The redshift we derive from $\textrm{C}~\textsc{iii}]$ is fully consistent with the redshift of Oxygen lines 4959--5007\AA\ identified in the MOSFIRE spectrum (see Table~\ref{tab:data}). This provides a robust estimate of the systemic redshift, $z=3.2127\pm0.0008$.

We derive ionization parameter, Oxygen and Carbon abundances using a modified version of the HII-CHI-mistry code \citep{perezmontero2014}, adapted to provide metallicity, C/O and ionization parameter in a Te-consistent framework, based on the comparison of the  observed UV and optical nebular lines with a grid of \textsc{cloudy} photoionization models \citep{ferlandetal2013}. The lines used as input are \civ, \ciii, \oiiii, \hb, while the \civ/\ciii\ (sensitive to the ionization parameter), \ciii/\oiiii\ (proportional to C/O), \oiiiv/\oiiii\ (sensitive to the electron temperature) and (\civ+\ciii)/\hb\ ratios are compared with model-based ratios in a $\chi^2$-weighted minimization scheme. The derived abundances and ionization parameter are $12+\log(\mathrm{O}/\mathrm{H})=8.07\pm0.44$, $\log(\mathrm{C}/\mathrm{O})=-0.80\pm0.13$, and $\log U=-2.25\pm0.81$. We also derive these quantities using all the lines except \oiiii\ ($S/N\sim1.7$) and we obtain $12+\log(\mathrm{O}/\mathrm{H})=7.79\pm0.35$ and $\log U=-2.32\pm0.11$ (C/O can not be derived without \oiiii). Both set of lines lead to low metallicity and high ionization parameter (in the following, we consider the results obtain with all the line measurements and upper limits). \ionii\ metallicity is similar to the typical green pea metallicity \citep[$12+\log(\mathrm{O}/\mathrm{H})=8.05\pm0.14$,][]{amorinetal2010}. The metallicity and ionization parameter are also consistent with extreme emission-line galaxies up to $z\sim3.5$ \citep{amorinetal2014,amorinetal2014b,amorinetal2015}.
The metallicity of \ionii\ is also consistent with the mean metallicity of star-forming galaxies selected through their extreme EW(\oiii) \citep{masedaetal2014,amorinetal2015,lyetal2014}.
Moreover, the \ionii\ C/O ratio is strongly subsolar \citep[$\mathrm{C}/\mathrm{O}_\odot =-0.26$][]{asplundetal2009} and consistent with the trend in C/O vs. O/H for local galaxies \citep{garnettetal1999}, which shows a nearly linear  increase in C/O with $Z>\sim0.2Z_\odot$. The low O/H and C/O abundances of \ionii\ are similar to those found in other strong emission-line galaxies at $z\sim1-3$ \citep[e.g.,][]{christensenetal2012,starketal2014b,jamesetal2014} and suggest a  significant contribution of O and C produced by massive stars, which is also consistent with very young and extreme starbursts.

\begin{figure}[htbf]
  \centering
  \includegraphics[width=9.5cm,trim=0cm 0cm 0cm 0cm,clip=true]{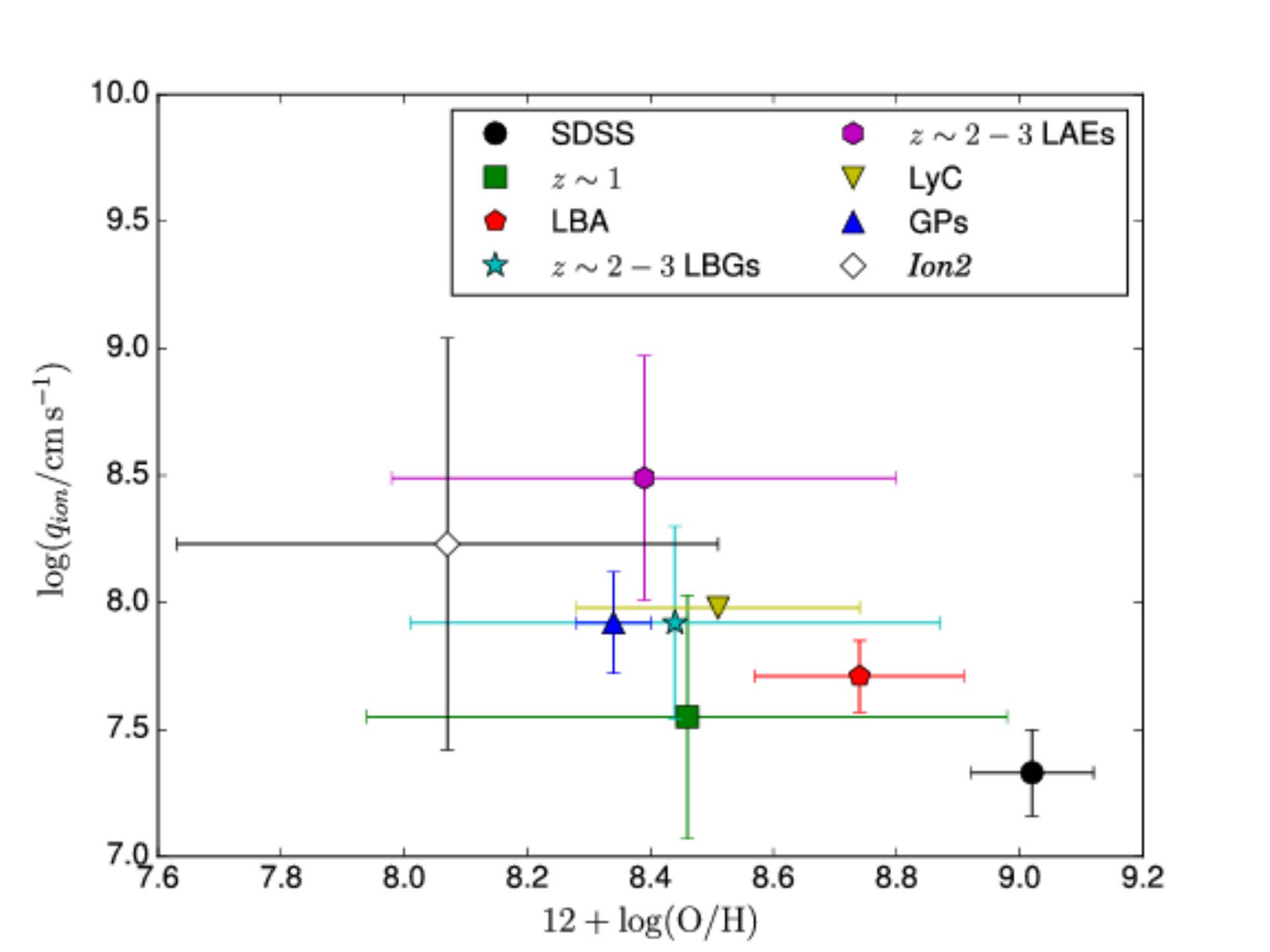}
    \caption{Comparison of the metallicity and ionization parameter ($U=q/c$) between the mean sample properties from \cite{nakajima&ouchi2014} and \ionii\ with 1$\sigma$ uncertainties. \ionii\ is compared to local galaxies from the SDSS survey, a galaxy sample at $z\sim1$, a Lyman break galaxy (LBG) sample at $z\sim2.5$, a Lyman alpha emitter (LAE) sample at $z\sim2.5$, local Lyman break analogues (LBAs), LyC emitters, and Green Peas (GPs). See \cite{nakajima&ouchi2014} and references therein for a complete description of these samples.}
  \label{fig:fig7}
\end{figure}

We compare the metallicity and the ionization parameter with the results presented in \cite{nakajima&ouchi2014} in Figure~\ref{fig:fig7}. Overall, \ionii\ has a lower metallicity than all other galaxy populations presented in their work and one of the highest ionization parameter. The higher ionization parameter is in line with a possible Lyman continuum leaking that could be explained with a low neutral hydrogen column density. Also, the \ionii\ extreme \oiiiv/\oii\ ratio, the metallicity, and the ionization parameter are consistent with \textsc{cloudy} models with a non zero Lyman continuum escape fraction \citep[][, Figure~11]{nakajima&ouchi2014}

The identification of Lyman continuum leakage from Green pea galaxies is a current line of research \citep[e.g.,][]{jaskot&oey2014,nakajima&ouchi2014,borthakuretal2014,yangetal2015} and the Green pea nature of \ionii\ and its LyC leakage represent the first concrete attempt to link these two properties. \ionii\ represents an extreme case of Green pea galaxy, being the highest redshift ($z>3$) ultra-strong Oxygen emitter (with $\mathrm{EW(\oiiidoub)}\sim1100$\AA) currently known.  The Lyman continuum leakage observed in \ionii\ allow us to investigate the relationship between the LyC leakage and physical and morphological properties. However, as shown in \cite{stasinskaetal2015}, the \oiii/\oii\ ratio is also related to the specific star formation rate and the metallicity: the \oiii/\oii\ ratio increases with increasing sSFR and decreasing metallicity. For \ionii\, the observed ratio is $\geq15$ making \ionii\ as an outlier in the Oxygen abundance vs. \oiii/\oii\ ratio relation and the EW(\hb) (i.e., sSFR) vs. \oiii/\oii\ ratio \citep[Figure 2,][]{stasinskaetal2015}: \ionii\ ratio is higher by $\sim0.6$ dex for the observed EW(\hb) (Table~\ref{tab:data}) and the ratio is higher by $\sim1$ dex compared to galaxies in the SDSS sample with similar metallicities. The \ionii\ \oiii/\oii\ ratio is also higher than what is expected for the derived stellar mass, SFR, and sSFR \citep{nakajima&ouchi2014}. Therefore, we conclude that the extreme \oiii/\oii\ ratio is due to unusual physical conditions (density-bounded nebula), which imply a low column density of neutral gas, and so favor leakage of ionizing photons \citep{nakajima&ouchi2014}.


\subsection{Physical properties derived from the photometry}
\label{sec:photprop}

\begin{figure}[htbf]
  \centering
  \includegraphics[width=9.5cm,trim=0cm 0cm 0cm 0cm,clip=true]{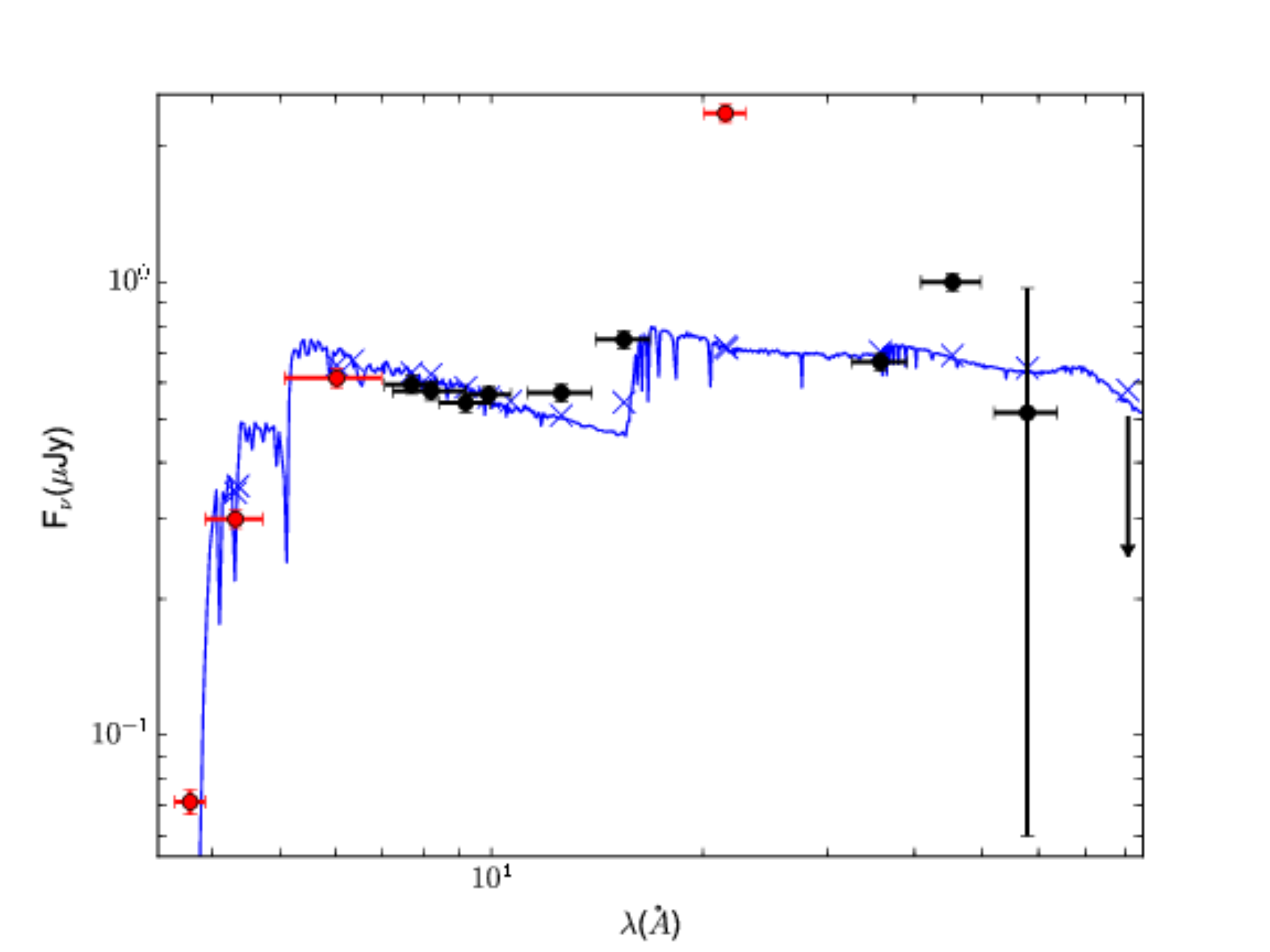}
    \caption{Observed \ionii\ SED (black and red dots). We show the best SED fit in blue and the integrated fluxes with blue crosses. We exclude photometric points (in red) affected by the IGM and strong emission lines (\lya, \oiii) from the SED fitting. The arrow indicates 1$\sigma$ upper limit.}
  \label{fig:fig8}
\end{figure}

While the emission lines highlight \ionii\ ISM physical properties, we also use the photometry to confirm findings such as the large \oiii\ emission and derive other physical parameters. From Figure~\ref{fig:fig8}, the jump in the K-band is evident,
$K=22.97 \pm0.02$ while the continuum level in the adjacent
bands, $H_{160}$ and IRAC Channel 1 is $\simeq 24.15$.
The $\Delta m \simeq 1.1$ magnitude in the HAWKI K-band
corresponds to an additional flux of $3.1\times10^{-16}$ erg/s/cm$^2$, that corresponds
to an observed equivalent width of 6700\AA\ (1600\AA~rest-frame). 
The total flux derived for \hb+\oiiidoub\ from the
MOSFIRE spectrum (see Tab.~\ref{tab:data})
is fully compatible with the flux
estimated from the $K$-band excess
(after correcting for slit losses of the three lines, see above).
The derived $H\beta$+[OIII] equivalent width is one of the largest ever observed
at $z>2$. The measured line ratios among the \hb+\oiiidoub\ suggest
the \oiiiv\ rest-frame equivalent width is $\simeq 1100\AA$, with a similar value to the few extreme
cases reported in literature at $z\sim1-2$ \citep{ateketal2011,amorinetal2014,masedaetal2014}. The \oiii/\hb\ ratio is also higher than the one reported in \cite{holdenetal2014} for LBGs at $z\sim3.5$, while the ratio is higher in this sample than in the local Universe \citep{brinchmannetal2004}.
Also at $z>6$ large equivalent widths of \hb+\oiii\ have been
reported, measured through photometric excess detected in the Spitzer/IRAC
channels \citep[e.g,][]{finkelsteinetal2013,smitetal2014,oeschetal2015}. While these cases can be spectroscopically investigated only with future
facilities,  like the {\it James Webb Space Telescope} ({\it JWST}) and Extremely Large Telescope class, in our case the MOSFIRE near infrared
spectroscopy has allowed us to access the details of the lines.

We derive the SFR from the UV luminosity, \hb\ and \lya\ lines \citep{kennicutt1998}, neglecting dust attenuation (see Appendix~\ref{ap:beta}). We get $\mathrm{SFR(UV)}=15.6\pm1.5\msunyr$, $\mathrm{SFR(\hb)}=40.0^{+7.5}_{-25.6}\msunyr$, and $\mathrm{SFR(\lya)}=14.3\pm1.0\msunyr$. The three SFR indicators are consistent within 68\% confidence levels, while SFR(\hb) is higher than the two others. The resonant nature of \lya\ photons prevents us from deriving precisely the SFR with this line \citep[e.g.,][]{ateketal2014b}. On the other hand, the \hb\ line is barely detected (Table~\ref{tab:data}). The higher SFR derived from the \hb\ line can be explained if the age of the galaxy is significantly younger than 100Myr because of the different equilibrium timescales between UV and nebular emission lines. However, this is not the case according to the SED fitting, with an age for \ionii\ $\sim200\mathrm{Myr}$. By comparing SFR(UV) and SFR(\lya), we derive a Lyman-$\alpha$ escape fraction $f_{esc}(\lya)\geq0.78$ ($f_{esc}(\lya)\geq0.28$ by using SFR(\hb) instead of SFR(UV)).
This high \lya\ escape fraction can be related to a blue $\beta$ slope \citep[e.g.,][]{hayesetal2014}, but also to a low \hi\ column density \citep[$N_{\hi}\leq10^{18}$ cm$^{-2}$;][]{yangetal2015,henryetal2015}. 

\ionii\ \lya\ profile has been described in V15: it exhibits a blue asymmetric tail. A low column density would be consistent with the bluer \lya\ emission at the systemic redshift, and also with the high ionization parameter, and so with a leakage of ionizing photons. The star formation activity confined in a small physical volume can also favor the escape of ionizing radiation \citep[e.g.,][]{heckmanetal2011,borthakuretal2014}. Furthermore, 
detailed analysis of the local blue compact dwarf galaxy NGC 1705 shows that O stars are able  to photoionize volumes far from their location \citep{annibalietal2015}, which could create density-bounded conditions.
Moreover, the 
spatial density of O stars in NGC 1705 increases 
toward the center of the galaxy, forming a sort of super star cluster efficient in ionizing its environment.
\ionii\ could resemble a similar condition, being spatially resolved, but nucleated in the ultraviolet (1400\AA), in which
a young super star cluster could be present in the core and significantly contributing 
(or totally contributing if the AGN is absent) 
to the ionization field.
Interestingly, because  \ionii\ has two components likely at a similar redshift (Section~\ref{sec:prev}), the leakage of ionizing photons can also be favored by the merging/interaction of these two components \citep{rauchetal2011}.

From SED fitting, we also derive a stellar mass $\log(\mstar/\msun)=9.5\pm0.2$ and an age of the stellar population (since the onset of the star formation) $\log(\mathrm{Age}/\mathrm{yr})=8.6\pm0.2$. We notice that while the two \ionii\ components are separated in the $BViz$ bands, they are seen as a unique galaxy in the other bands, and
the stellar mass is the total stellar mass of the two components.
Given the very close separation among the two (0.2''), we decided to
avoid any tentative decomposition and SED fitting of each
component individually, especially in the Spitzer/IRAC bands where
the pixel size is 1.22'' and the decomposition is in practice
not possible.
Using SED fitting and anchoring the best fit to the $\beta$ slope derived from UV spectrum fitting (Appendix~\ref{ap:beta}), we predict the expected observed L(IR)$_{24\mu\mathrm{m}}$ assuming energy balance (i.e, we assume a balance between the energy absorbed in the UV/optical (we integrate over 912\AA\ to 3\micron) and re-radiated in the IR). The MIPS 24\micron\ observations probe the PAH feature at 6.2\micron\ (rest-frame) which can be used to derive the total IR luminosity (i.e., IR luminosity integrated over 8-1000\,\micron) and then the bolometric SFR \citep{popeetal2008,rujopakarnetal2012}. Because the SFR derivation relies on several assumption \citep[star formation history, age, metallicity, IMF,][]{kennicutt1998}, we rely on the comparison between predicted L(IR)$_\mathrm{SED}$ and derived L(IR)$_{24\mu\mathrm{m}}$, rather than relying on a SFR comparison.  From the MIPS 24\micron\ upper limit, we derive $\log(\mathrm{L(IR)/\lsun})_{24\mu\mathrm{m}}=9.92^{+0.65}_{-0.44}$, while from SED fitting we derive $\log(\mathrm{L(IR)/\lsun})_\mathrm{SED}=9.77^{+1.00}_{-2.03}$. The SED predicted and observed L(IR)  are consistent but the uncertainties are too large to exclude an AGN contribution to the IR flux.
We further discuss a possible AGN contribution in Section~\ref{sec:agn}.

\ionii\ also shows an excess in the IRAC2 band, in comparison with the IRAC1 channel. This excess can not be explained by emission line or other feature from both blobs (e.g., a contamination by a strong emission line as \ha\ would imply $z>5.0$ for the emitting galaxy). While we do not have a satisfactory explanation, we note that the rest-frame wavelength interval probed by IRAC2 (9500\AA-11900\AA) cannot be explored in detail before the launch of  {\it JWST}. Therefore that part of the spectrum is poorly known, especially for sources in the distant universe. In this work we do now comment further about this feature that, however,  do not affect our results.
We summarize the physical properties of \ionii\ in Table~\ref{tab:param}.

      \begin{figure}[htbf]
  \centering
  \includegraphics[width=8.5cm,trim=0cm 0cm 0cm 0cm,clip=true]{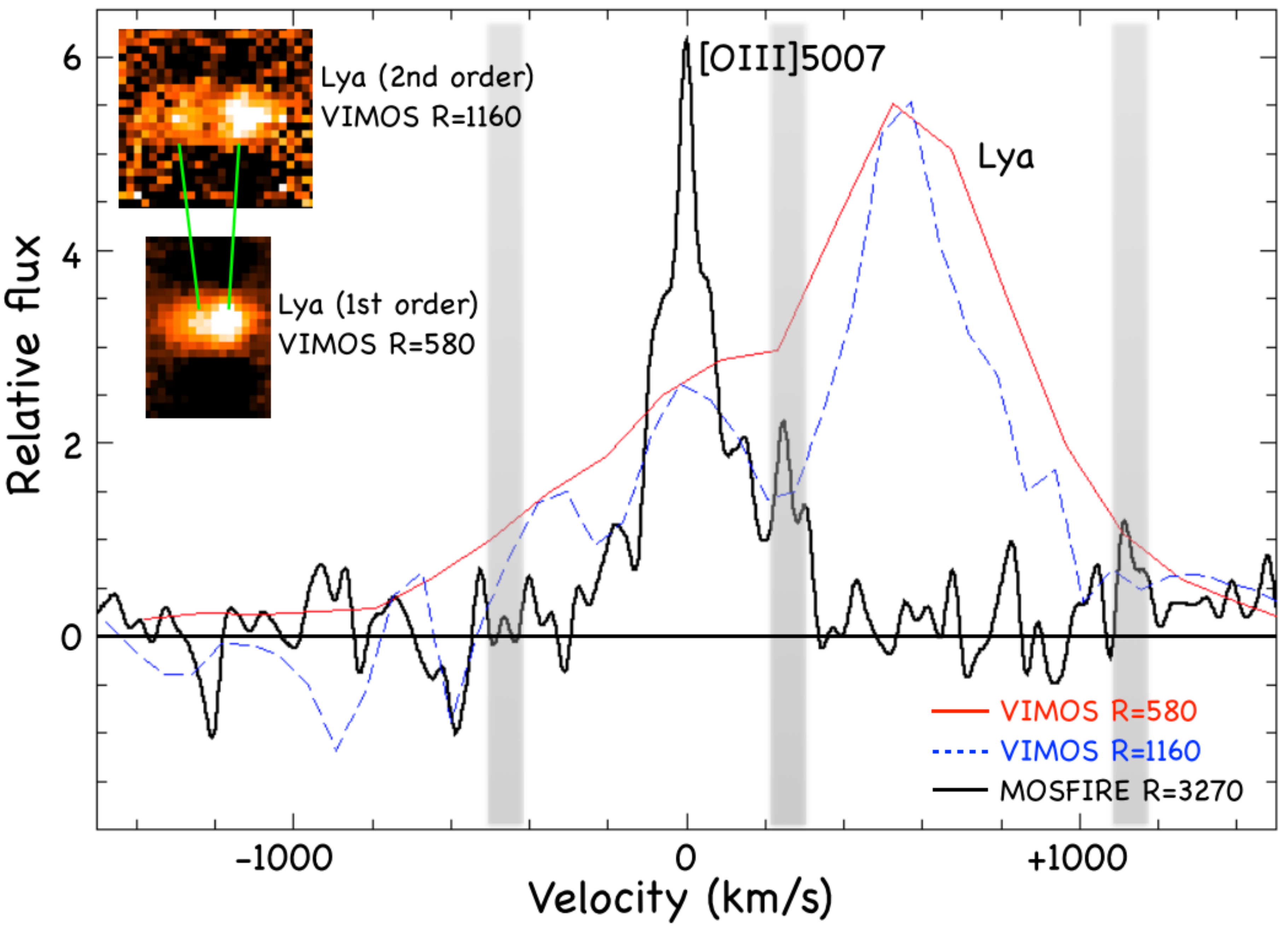}
    \caption{The velocity profiles of \oiiiv\ and \lya\ lines relative to systemic. \oiiiv\ line is shown in black, the \lya\ first order in red, and the \lya\ second order with the dashed blue line. The positions and width of the sky lines are marked with gray stripes.}
  \label{fig:fig9}
\end{figure}

        \begin{table}[htbf]
         \centering
         \caption{\ionii\ physical properties.}
         \begin{tabular}{lc}
         \hline
         $\log (\mstar/\msun)$ & $9.2\pm0.2$ \\
         $\log (\mathrm{age}/\mathrm{yr})$ & $8.6\pm0.2$ \\
         SFR (\msunyr) & $15.6\pm1.5$ \\
         $\mathrm{E(B-V)}_\mathrm{stars}$ & $\leq0.04$ \\
         $12+\log(\mathrm{O}/\mathrm{H})$ & $8.07\pm0.44$ \\
         $\log q$ & $8.23\pm0.81$ \\
         $\log(\mathrm{C}/\mathrm{O})$ & $-0.80\pm0.13$ \\
           \hline
         \end{tabular}
         \label{tab:param}
      \end{table}


  \subsection{Properties previously reported: summary}
  \label{sec:prev}
  
  \ionii\ has been previously reported as a Lyman continuum emitter candidate in V15. Here, we briefly summarize the \ionii\ properties reported in this latter reference.
  
\ionii\ is a compact source ($300\pm70$ parsecs) with two visible components. The presence of these two blobs can be misleading when trying to assess the detection of 
a possible Lyman continuum emission, because multiple components are often source of foreground contamination \citep{vanzellaetal2010b,sianaetal2015,mostardietal2015}. However, $BViz$ magnitudes have been  derived for each component after subtracting the close companion with a GALFIT fitting procedure
(see V15). The two show the same break between the $V_{606}$ and $B_{435}$ bands and the absence of any spurious (low-$z$) emission line along the wide wavelength range probed here (from 3400\AA\ up to 24000\AA) support that the faint component is a physical companion of the brightest one (i.e., at $z=3.2$). Furthermore, the slit orientation in the MOSFIRE observation is also maximizing the separation in the spatial direction (see Figure~\ref{fig:fig1}), though seeing FWHM is significantly larger than the separation. Therefore from photometric shape and spectroscopic arguments we conclude that the two components
are at the same redshift.

Interestingly, the \lya\ profile exhibits a double peaked emission, with the bluer component being half of the main red peak and having a redshift consistent with the systemic redshift ($z=3.212$, Figure~\ref{fig:fig9}). If the two \lya\ components are due to the two \ionii\ blobs, we would expect multiple-peaked \oiii\ emissions \citep{kulasetal2012}, which are not present in the \ionii\ MOSFIRE spectrum (see Section~\ref{sec:elprop}). If the observed \lya\ photons arise only from the brightest \ionii\ component, as discussed above it can be consistent with a low neutral gas column density  \citep[e.g.,][]{verhammeetal2015}, and with the escape of Lyman continuum photons \citep{behrensetal2014}. However, a small shift of the maximum of the profile compare to the systemic redshift is expected in case of Lyman continuum emission (i.e., $\log(\mathrm{N}_{\hi})<18$), with a maximum peak separation $<300$km s$^{-1}$ \citep{verhammeetal2015}. But the non detection of low-ionization absorption lines (e.g., \cii, \siii$\lambda1260$) is consistent with a low neutral hydrogen column density allowing the escape of ionizing photons and a low covering fraction along the line of sight \citep{heckmanetal2011,jonesetal2012,jonesetal2013}.  We possibly detect
 $\textrm{O}~\textsc{i}+[\textrm{Si}~\textsc{ii}]\lambda1303$ with $\mathrm{EW}=-1.2\pm0.7\AA$\ and tentatively $[\textrm{Si}~\textsc{ii}]\lambda1526$ with $\mathrm{EW}=-1.0\pm0.7\AA$. However, the signal-to-noise ratio is low ($S/N\sim2$). Assuming a 1$\sigma$ fluctuation, we can put a lower limit on the EW of non detected lines as $\mathrm{EW}>-0.7\AA$. 
This is in line with the results from \cite{shapleyetal2003} at $z\sim3$: galaxies with high \lya\ equivalent width exhibit much weaker low-ionization absorption lines than galaxies with low EW(\lya).

Finally, \ionii\ is not detected in the 6Ms X-ray Chandra image ($L_X<2-3\times10^{42}\mathrm{erg\hspace{1mm}s}^{-1}$) and most of the high-ionization emission lines are not detected (e.g., \nv, \civ). However, from the UV spectrum reanalysis, we report a tentative detection of \heii\ (see Sect.~\ref{sec:elprop}): this line is is barely detected in
the ABBA spectrum, but not in the \cite{balestraetal2010} spectrum.
 Furthermore, V15 also reported a possible variability of \ionii. We further discuss the possibility of an AGN contribution in Section \ref{sec:agn}.


 \section{Is \ionii\ an AGN?}
  \label{sec:agn}
  
We now discuss the possibility that the main source of UV ionizing photons is an AGN.  
First of all, the UV emission is resolved so it is most probably due to the stellar population, not an AGN. Therefore, an AGN could contribute to the SED but it would not dominate it at all wavelengths.

Unfortunately, we can not use the BPT diagram \citep{baldwinetal1981} to discriminate between star-formation and AGN, because \ha\ and \nii\ emission lines are out of the MOSFIRE wavelength range. \cite{juneauetal2011} proposed an alternative to the BPT diagram, with the so-called MEx diagram based on the comparison between stellar mass and \oiiiv/\hb\ ratio. The probability for an object to be an AGN associated with the MEx diagram is not defined at the \ionii\ position \citep{juneauetal2014}, although the large \oiiiv\ to \hb\ ratio would classify \ionii\ as an AGN. Other diagnostic diagrams have been proposed, like the \oiiiv/\hb\ vs. \oiilam/\hb\ \citep{lamareilleetal2004}. Also in this diagram, \ionii\ would be classified as an AGN, because of the large \oiiiv/\hb\ ratio. However, for both diagnostics (MEx and \oiiiv/\hb\ vs. \oiilam/\hb), there is no known AGN at a similar locus as \ionii. Alternatively, photoionization models predict a possible increase of the \oiii/\oii\  with decreasing metallicity and increasing ionization parameter \citep{nakajima&ouchi2014}.
  
As already stated, \ionii\ remains undetected even in the 6Ms X-ray Chandra survey. 
According to the empirical relation between the \oiiiv\ luminosity and the X-ray luminosity \citep[e.g.,][]{panessaetal2006},
given the extreme \oiiiv\ emission 
the expected X-ray luminosity would be
$L_X\sim10^{45}\mathrm{erg\hspace{1mm}s}^{-1}$ if the \oiiiv\ emission is associated to an AGN. Considering the scatter in the $L(\oiiiv)$-$L_X$ relation, we can derive a lower limit of $L_X\geq5\times10^{43}\ergs$ which would be easily detectable in the 6Ms Chandra data (with typical $S/N > 10$), independently on the source position in the X-ray image. 
However, extremely obscured AGN can have lower X-ray luminosity than expected from the $L(\oiiiv)$-$L_X$ relation, this AGN possibly being Compton-thick AGN \citep[e.g.,][]{vignalietal2010}. The infrared and X-ray luminosities derived from the MIPS detection and the 6Ms upper limit, respectively, can be compared with the properties of known  obscured AGN \citep{lanzuisietal2015}: \ionii\ properties can be only consistent with a faint and highly obscured AGN \citep[e.g.,][]{lutzetal2004}.

The large EW(\oiiiv) is unusually large for all types of AGN \citep[Table~\ref{tab:data},][]{caccianiga&severgnini2011}, while such large equivalent width has been already reported for some star-forming galaxies at lower redshift \citep[e.g.,][]{ateketal2011,vanderweletal2011,brammeretal2012b}.

The UV spectrum is quite unexplored as a diagnostic for
AGN/star-forming galaxies classification, but the observed narrow lines (Table~\ref{tab:data}) exclude the presence of broad-line AGN. We also compare \ionii\ emission line ratios with typical UV emission line ratios in narrow line AGN. \cite{nagaoetal2006} present the UV spectrum analysis of narrow-line QSOs and narrow-line radio galaxies at $1.2\leq z\leq3.8$ \citep[see also][]{debreucketal2000}. The first evidence against a possible AGN contribution to \ionii\ spectra is the lack of \civ\ detection.
This line is detected in all AGN types studied in \cite{nagaoetal2006}. The typical line ratio \ciii/\civ\ is $\sim0.5$ and $\civ/\heii\geq1$, while we measure $>10$ and $<1$ for \ionii\, respectively. Narrow-line AGN UV spectra at $z\sim2-3$ exhibit a \ciii/\civ\ ratio near 1 which is also much lower than the \ionii\ ratio \citep{hainlineetal2011}. 
The \civ/\ciii\ ratio can be used to determine the nature of the ionizing source, with  $\civ/\ciii\sim2$ if the ionizing photons are produced by an AGN and $\civ/\ciii\sim0.001$ if population I stars are the ionizing sources \citep{binetteetal2003}. The clear \ciii\ detection and the \civ\ non detection are consistent with a main ionizing source being stars rather than an AGN. 

The origin of \heii\ emission in galaxies has been matter of investigation, possibly produced by massive stars both in the local Universe \citep[e.g.,][]{szecsietal2015,grafener&vink2015} and at high-redshift \citep[e.g.,][]{cassataetal2013,sobraletal2015,pallottinietal2015}.
\heii\ detection is generally associated with AGN activity, but it has also been reported that \heii\ associated with no \civ\ is a possible feature in star-forming galaxies at $z\sim3$ \citep{cassataetal2013}. This could be explained by a contribution from population III stars \citep{sobraletal2015}. However, as stated in the previous section,
 we remind that the \heii\ detection is tentative.

Summarizing, \ionii\ is not detected in the 6Ms Chandra survey and has no broad lines. The extreme \oiii/\hb\ ratio would lead to consider it as an AGN \citep{lamareilleetal2004,juneauetal2014}, but this ratio can also be explained by extreme ISM physical conditions \citep{jaskot&oey2013,nakajima&ouchi2014}. The lack of X-ray detection can be explained with an obscured AGN \citep{vignalietal2010}, but the lack of \civ\ detection in the UV spectrum is likely inconsistent with this hypothesis. Therefore, the main argument against an AGN contribution is the lack of the \civ\ emission line. If an AGN contributes to the \ionii\ emission, it is likely a faint and obscured AGN with peculiar photoionization conditions \citep{debreucketal2000}.
    

 \section{Conclusions}
 \label{sec:conclu}
 In this paper, we present  new observations with the Keck/MOSFIRE NIR spectrograph and a new analysis of the UV spectrum of a Lyman continuum emitter candidate. \ionii\ is an object at $z=3.212$ composed of two distinct blobs, nucleated and resolved in the rest-frame UV, indicating  emission from star-forming regions. \ionii\ is also a candidate Lyman continuum emitter with $U$ band flux consistent with a leakage of ionizing photon ($P(\flyc=0)<10^{-4}$), as reported in V15. Contamination from foreground galaxies can be ruled out.
 
Our main results can be summarized as follows: 
\begin{itemize}
\item a new analysis of the UV spectrum shows a signal consistent with a direct detection of ionizing photons with $S/N>5$ and a  relative Lyman continuum escape fraction $f_{esc,rel}=0.64^{+1.10}_{-0.10}$;
\item the \lya\ emission at the systemic redshift, the high \lya\ escape fraction, the non detection of low-ionization absorption lines are consistent with a low neutral hydrogen column density, while velocity separation of the two \lya\ peaks is in tension with expectation \cite[e.g.,][]{verhammeetal2015};
\item we find low metallicity ($\sim1/6Z_\odot$), strongly subsolar C/O ratio and high ionization parameter ($\log U=-2.25$) using a $T_e$-consistent method, in good agreement with previous results at $z\sim2-3$;
\item \ionii\ exhibit one of the largest \oiii/\oii\ ratio observed at $z>3$ and similar large ratios are predicted for galaxies with low metallicities and Lyman continuum leakage \citep{nakajima&ouchi2014};
\item there is no clear evidence of AGN contribution to \ionii\ emission;
\item the \ionii\ physical properties (SFR, stellar mass, dust attenuation, strong emission lines) are similar to green peas, and the \ionii\ large EW(\hb+\oiiidoub) is similar with the ones derived for LBGs at $z>7$.
\end{itemize}

\ionii\ physical properties, morphology, and very strong emission lines make it a possible analog of $z\sim8$ LBGs. Furthermore, unlike local analogs \citep[e.g.,][]{heckmanetal2011}, \ionii\ probes Lyman continuum emitter properties only $\sim1.5$Gyr after the reionization epoch. From the very large EW(\oiiidoub+\hb) observed at high-redshift \citep{oeschetal2015,zitrinetal2015,robertsborsanietal2015,smitetal2015}, we can speculate that these galaxies have a leakage of ionizing photons.
 
The direct detection of Lyman continuum emission leave little doubt about the fact that \ionii\ emits ionizing flux.
This galaxy certainly exhibits peculiar properties which however do not allow to fully discriminate among different possible sources of ionizing radiation yet: stellar emission, faint and obscured AGN, or other ionizing sources. Only high resolution {\it HST} observations of the ionizing radiation may provide useful clues. In the near future, our approved proposal to
observe \ionii\ with {\it HST}/F336W will hopefully shed new light on the nature of this source (PI: Vanzella).

\begin{acknowledgements}
We thank the referee for suggestions for clarifying the text and the analysis.
We thank M. Tosi, F. Annibali, M. Brusa and H. Atek for useful discussions.
We acknowledge the financial contribution from PRIN-INAF 2012. 
This work has made use of the Rainbow Cosmological Surveys Database, which is operated by the Universidad Complutense de Madrid (UCM), partnered with the University of California Observatories at Santa Cruz (UCO/Lick, UCSC).
\end{acknowledgements}

\bibliographystyle{aa}
\bibliography{ref}

\appendix

\section{Emission lines dust correction}
\label{ap:el}

Emission lines are often dust corrected using the Balmer decrement (\ha/\hb): the observed ratio between Balmer lines is compared with the expected ratio derived from theory \citep[e.g., $\ha/\hb=2.86$;][]{osterbrock1989}, and assuming an attenuation curve, the dust correction is derived at any wavelength \citep[e.g.,][]{kashinoetal2013,dominguezetal2013,priceetal2014,steideletal2014}. Unfortunately, regarding \ionii, the \ha\ line is outside the MOSFIRE wavelength range and we must rely on the dust attenuation derived from the UV $\beta$ slope \citep[$f_\lambda\propto\lambda^\beta$;][]{meureretal1999} or from SED fitting (which basically also relies on the fit of the UV $\beta$ slope).

\cite{calzettietal2000} derive a relation between the stellar and nebular color excess as,
\begin{eqnarray}
\label{eq:color}
\mathrm{E(B-V)}_\mathrm{nebular}=2.27\times\mathrm{E(B-V)}_\mathrm{stellar},
\end{eqnarray}
and this relation is often misunderstood as both color excesses being derived with the same attenuation curve. Actually the nebular color excess is derived from a line-of-sight attenuation curve \citep[e.g.,][]{cardellietal1989}, while the stellar continuum color excess is derived with the Calzetti curve \citep{calzettietal2000}.

This issue is challenged by contradictory result about the stellar color excess relative to the nebular color excess at $z\sim2$: some studies found results consistent with the misinterpretation described previously \citep[e.g.,][]{FS09}, while other found $\mathrm{E(B-V)}_\mathrm{nebular}\sim\mathrm{E(B-V)})_\mathrm{stellar}$ \citep[e.g.,][]{erbetal2006b,reddyetal2010,shivaeietal2015}. A detailed comparison of a large sample of $z\sim2$ star-forming galaxies observed with MOSFIRE and with measured Balmer decrement for most of this sample \citep{reddyetal2015}, shows that the relation between $\mathrm{E(B-V)}_\mathrm{nebular}$ and $\mathrm{E(B-V)}_\mathrm{stellar}$ is in fact SFR (and sSFR) dependent \citep[see also][]{debarrosetal2015}. To derive the emission line attenuation, we use the relation provided by \cite{reddyetal2015}:
\begin{eqnarray}
\label{eq:attdiff}
\mathrm{E(B-V)}_\mathrm{nebular}-\mathrm{E(B-V)}_\mathrm{stellar}=-0.049+0.079\nonumber\\\times(\log(\mathrm{sSFR}_\mathrm{SED})+10)
\end{eqnarray}
\cite{reddyetal2015} emphasize that Equation~\ref{eq:attdiff} is affected by a large scatter ($\sigma\sim0.12$).


\section{UV $\beta$ slope and SED fitting procedure}
\label{ap:beta}

 \begin{figure}[htbf]
  \centering
  \includegraphics[width=9.5cm,trim=0cm 0cm 0cm 1.3cm,clip=true]{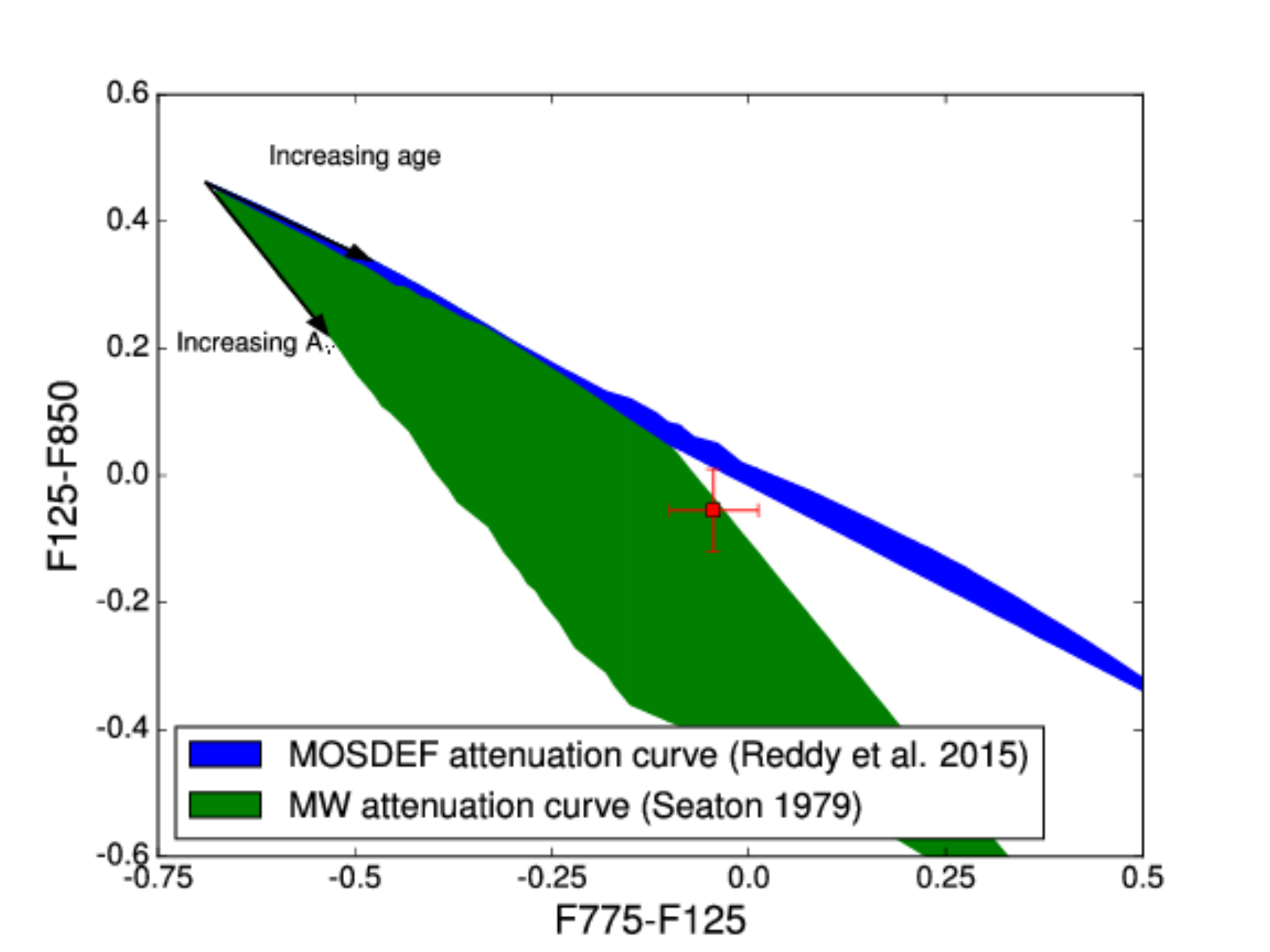}
    \caption{Range of $\mathrm{F775}-\mathrm{F125}$ and $\mathrm{F125}-\mathrm{F850}$ colors predicted with BC03 templates and the MOSDEF \citep{reddyetal2015} and MW attenuation curves \citep{seaton1979}. \ionii\ colors are shown with the red square.}
  \label{fig:dustcurve}
\end{figure}

 \begin{figure}[htbf]
  \centering
  \includegraphics[width=9cm,trim=0cm 0cm 0cm 0cm,clip=true]{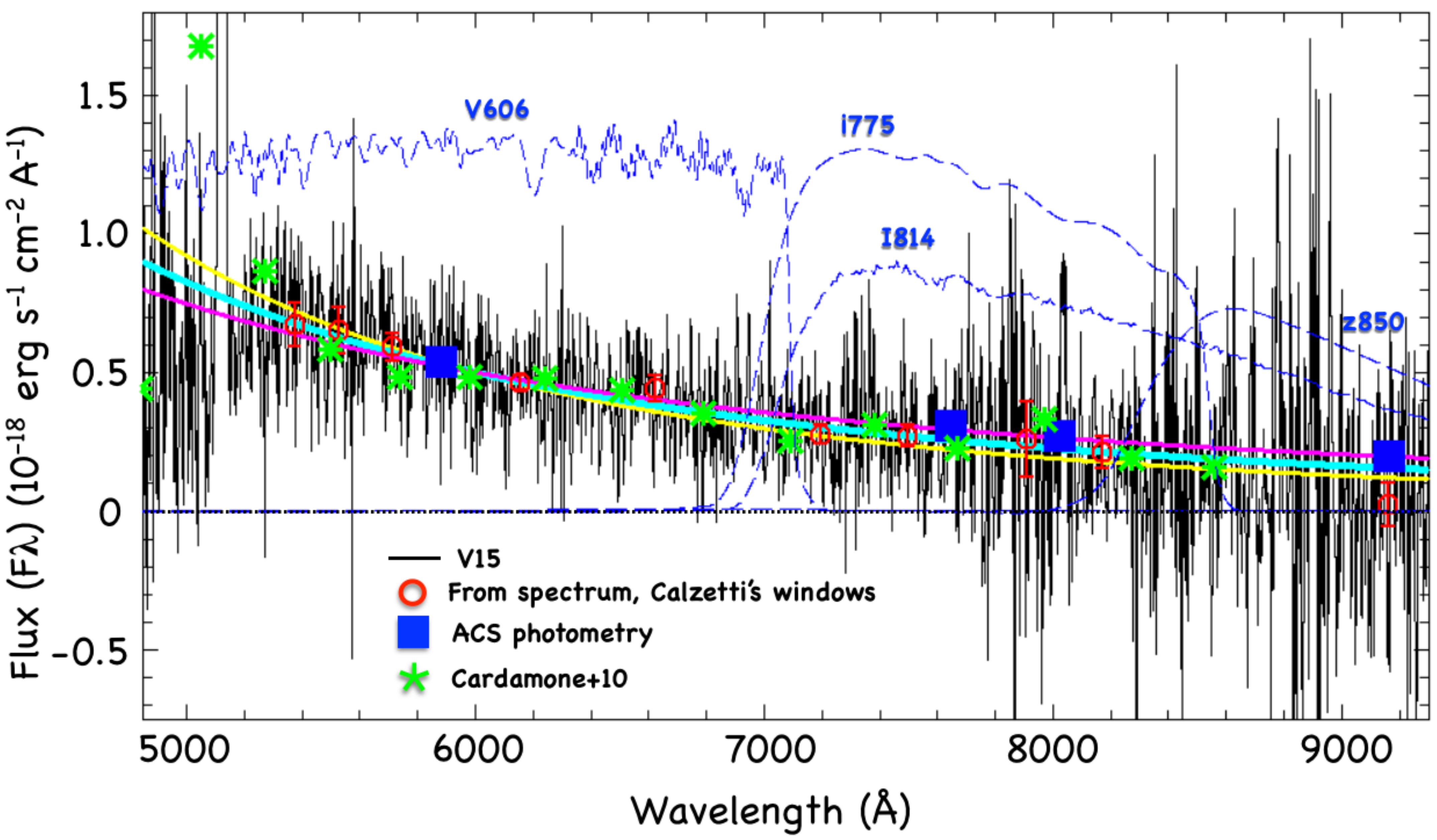}
    \caption{1D UV spectrum from V15 (black). ACS photometry (blue squares) is shown along with the corresponding transmission curves ($V_{606}$, $i_{775}$, $I_{814}$, and $z_{850}$, blue dashed lines). We also show the medium band photometry from the MUSYC survey \citep[green stars,][]{cardamoneetal2010}. The red dots and associated error bars show the flux derived through the wavelength windows used in \cite{calzettietal1994} to derive the UV $\beta$ slope.  The cyan line denotes the best fit over the Calzetti windows and the violet and yellow lines encompass the 68\% uncertainty ($\beta =-2.75$, -3.40 and -2.20, respectively).}
  \label{fig:betauv}
\end{figure}

We want to derive the UV $\beta$ slope as accurately as possible because this quantity is known to be related to the dust attenuation, while also dependent on the age of the stellar population and the star formation history \citep{leitherer&heckman1995,meureretal1999}. To derive the UV $\beta$ slope, we perform a multi-band fitting, using the bands between 1500\AA\ and 2500\AA\ \citep{castellanoetal2012}. However, the flux in the $z_{850}$ band is lower than what we expect from the observed fluxes in the band just blueward and redward $z_{850}$ ($I_{814}$ and $Y_{098}$).
We compare theoretical expectation from the MOSDEF attenuation curve and the MW curve \citep{seaton1979} with \ionii\ observed colors in Figure~\ref{fig:dustcurve}. The MW curve exhibits the 2175\AA\ bump which could explain the flux decrease in the $z_{850}$ band, while uncertainties are too large to discriminate between a curve with or without the 2175\AA\ bump. However, \cite{scovilleetal2015} report that their data are consistent with the presence of the 2175\AA\ bump, while for \cite{reddyetal2015} this presence is only marginal.

We derive the $\beta$ slope from the UV spectrum, using the wavelength windows described in \cite{calzettietal1994}, but excluding the 10th window because being outside the UV spectrum wavelength coverage. The median flux within each window is calculated and used to fit a linear relation in the observed spectrum, assuming the flux density per unit wavelength $F_\lambda\sim\lambda^{\beta}$, in units of erg s$^{-1}$ cm$^{-2}$\AA$^{-1}$\ \citep{hathietal2015}. The resulting slope is $\beta=-2.75 _{-0.65}^{+0.55}$. It is worth noting  that the UV slope from the spectrum is not influenced by the possible  2175\AA\ bump. The UV slope derived from the spectrum is consistent within 1$\sigma$ with the $\beta$ slope derived from multi-band fitting of the photometry ($-2.20\pm0.20$).

Regarding the SED fitting procedure, we use BC03 templates \citep{BC03} and we anchor the result to the UV $\beta$ slope by not taking into account solutions inconsistent with the $\beta$ slope derived from the UV spectrum, within 1$\sigma$. We also fix the metallicity to $Z=0.2Z\sun$, which is consistent with the metallicity derived in Section~\ref{sec:elprop}. We test different star formation histories (declining, rising, constant) and attenuation curves (Calzetti, MOSDEF, MW), assuming a Salpeter IMF \citep{salpeter1955}. The results show that the physical parameters such as the stellar mass and the age are not strongly affected by the assumptions. Therefore, in this work, we assume a constant star formation history and the MOSDEF attenuation curve. While the $\beta$ slope is consistent with no or low dust attenuation, we use the relations derived in \cite{reddyetal2015} to translate the $\beta$ slope in color excess. This leads to $\mathrm{E(B-V)}_\mathrm{stellar}\leq0.04$, while the Meurer relation \citep{meureretal1999} leads to no dust attenuation because of the higher intrinsic $\beta$ slope in this latter relation.


\end{document}